\documentclass[pre,floatfix,superscriptaddress,twocolumn,final]{revtex4-1}
\usepackage{amsmath}
\usepackage{amssymb}
\usepackage{amsthm}
\usepackage{scrextend}
\usepackage{graphicx} 
\usepackage{hyperref}
\hypersetup{
    bookmarks=true,         
    unicode=false,          
    pdftoolbar=true,        
    pdfmenubar=true,        
    pdffitwindow=false,     
    pdfstartview={FitH},    
    pdftitle={The ground truth about metadata and community detection in networks},    
    pdfauthor={Leto Peel, Daniel B. Larremore, Aaron Clauset},     
    pdfcreator={},   
    pdfproducer={}, 
    pdfkeywords={} {} {}, 
    pdfnewwindow=true,      
    colorlinks=true,       
    linkcolor=red,          
    citecolor=Green,        
    filecolor=magenta,      
    urlcolor=cyan           
}
\usepackage[usenames,dvipsnames]{color}
\definecolor{matrix}{rgb}{0.5,0.65,0.5}

\DeclareMathOperator*{\argmin}{arg\,min}

\begin{document}

\author{Leto Peel}
\thanks{Contributed equally}
\email{leto.peel@uclouvain.be}
\affiliation{ICTEAM, Universit\'{e} Catholique de Louvain, Louvain-la-Neuve, Belgium}
\affiliation{naXys, Universit\'{e} de Namur, Namur, Belgium}

\author{Daniel B. Larremore}\thanks{Contributed equally}
\email{larremore@santafe.edu}
\affiliation{Santa Fe Institute, Santa Fe, NM 87501, USA}

\author{Aaron Clauset}
\email{aaron.clauset@colorado.edu}
\affiliation{Department of Computer Science, University of Colorado, Boulder, CO 80309, USA}
\affiliation{BioFrontiers Institute, University of Colorado, Boulder, CO 80309, USA}
\affiliation{Santa Fe Institute, Santa Fe, NM 87501, USA}

\newcommand*\mycdot{{\mkern 2mu\cdot\mkern 2mu}} 	
\newcommand*\dif{\,\mathrm{d}} 					

\newcommand{\LP}[1]{\textcolor{red}{#1}}
\newcommand{\LPcom}[1]{[\textcolor{red}{LP: #1}]}

\newcommand{\DL}[1]{\textcolor{blue}{#1}}
\newcommand{\DLcom}[1]{[\textcolor{blue}{DL: #1}]}

\newcommand{\AC}[1]{\textcolor{purple}{#1}}
\newcommand{\ACcom}[1]{[\textcolor{purple}{AC: #1}]}

\newcommand{\matrixquote}[1]{\hrulefill \par \noindent \textcolor{matrix}{#1} \par \noindent \hrulefill \vspace{3mm}} 


\begin{abstract}
Across many scientific domains, there is a common need to automatically extract a simplified view or coarse-graining of how a complex system's components interact. This general task is called community detection in networks and is analogous to searching for clusters in independent vector data. It is common to evaluate the performance of community detection algorithms by their ability to find so-called \textit{ground truth} communities. This works well in synthetic networks with planted communities because such networks' links are formed explicitly based on those known communities. However, there are no planted communities in real world networks. Instead, it is standard practice to treat some observed discrete-valued node attributes, or metadata, as ground truth. Here, we show that metadata are not the same as ground truth, and that treating them as such induces severe theoretical and practical problems. We prove that no algorithm can uniquely solve community detection, and we prove a general No Free Lunch theorem for community detection, which implies that there can be no algorithm that is optimal for all possible community detection tasks. However, community detection remains a powerful tool and node metadata still have value so a careful exploration of their relationship with network structure can yield insights of genuine worth. We illustrate this point by introducing two statistical techniques that can quantify the relationship between metadata and community structure for a broad class of models. We demonstrate these techniques using both synthetic and real-world networks, and for multiple types of metadata and community structure.
\end{abstract}

\title{The ground truth about metadata and community detection in networks}

\maketitle


\section*{Introduction}
Community detection is a fundamental task of network science that seeks to describe the large-scale structure of a network by dividing its nodes into communities (also called blocks or groups), based only on the pattern of links. This task is similar to that of clustering vector data, as both seek to identify meaningful groups within some data set.

Community detection has been used productively in many applications, including identifying allegiances or personal interests in social networks~\cite{adamic2005political,fortunato2010community}%
, biological function in metabolic networks~\cite{holme2003subnetwork,guimera2005functional}, fraud in telecommunications networks~\cite{cortes2001communities}, and homology in genetic similarity networks~\cite{haggerty2014pluralistic}. Many approaches to community detection exist, spanning not just different algorithms and partitioning strategies, but also fundamentally different definitions of what it means to be a ``community.'' This diversity is a strength, as networks generated by different processes and phenomena should not \textit{a priori} be expected to be well-described by the same structural principles.

With so many different approaches to community detection available, it is natural to compare them to assess their relative strengths and weaknesses. Typically, this comparison is made by assessing a method's ability to identify so-called \textit{ground truth} communities, a single partition of the network's nodes into groups which is considered \textit{the} correct answer. This approach for evaluating community detection methods works well in artificially generated networks, whose links are explicitly placed according to those ground truth communities and a known data generating process.  For this reason, the partition of nodes into ground truth communities in synthetic networks is called a planted partition. However, for real-world networks, both the correct partition and the true data generating process are typically unknown, which necessarily implies that there can be no ground truth communities for real networks. Without access to the very thing these methods are intended to find, objective evaluation of their performance is difficult.

Instead, it has become standard practice to treat some observed data on the nodes of a network, which we call node \textit{metadata}, (e.g., a person's ethnicity, gender or affiliation for a social network, or a gene's functional class for a gene regulatory network) as if they were ground truth communities. While this widespread practice is convenient, it can lead to incorrect scientific conclusions under relatively common circumstances. In this paper, we identify these consequences and articulate the epistemological argument against treating metadata as ground truth communities. Next, we  provide rigorous mathematical arguments and prove two theorems that render the search for a universally best ground-truth recovery algorithm as fundamentally flawed. We then present two novel methods that can be used to productively explore the relationship between observed metadata and community structure, and demonstrate both methods on a variety of synthetic and real-world networks, using multiple community detection frameworks. Through these examples, we illustrate how a careful exploration of the relationship between metadata and community structure can shed light on the role that node attributes play in generating network links in real complex systems.

\section*{Results}
\subsection*{The trouble with metadata and community detection}

The use of node metadata as a proxy for ground truth stems from a reasonable need: since artificial networks may not be representative of naturally occurring networks, community detection methods must also be confronted with real-world examples to show that they work well in practice. If the detected communities correlate with the metadata, we may reasonably conclude that the metadata are involved in or depend upon the generation of the observed interactions.  However, the scientific value of a method is as much defined by the way it fails as by its ability to succeed. Because metadata always have an uncertain relationship with ground truth,
failure to find a good division that correlates with our metadata is a highly confounded outcome, arising for any of several reasons:
\begin{enumerate}
	\setlength\itemsep{0.01em}
	\item [(i)] these particular metadata are irrelevant to the structure of the network,
	\item [(ii)] the detected communities and the metadata capture different aspects of the network's structure, 
	\item [(iii)] the network contains no communities as in a simple random graph~\cite{erdHos1959random} or a network that is sufficiently sparse that its communities are not detectable~\cite{decelle2011prl}, or 
	\item [(iv)] the community detection algorithm performed poorly. 
\end{enumerate}
In the above, we refer to the \textit{observed} network and metadata and note that noise in either could lead to one of the reasons above.  For instance, measurement error of the network structure may make our observations unreliable and in extreme cases can obscure community structure entirely, resulting in case (iii). It is also possible that human errors are introduced when handling the data, exemplified by the widely used American college football network~\cite{girvan2002community} of teams that played each other in one season, whose associated metadata representing each team's conference assignment were collected during a different season~\cite{evans2010clique}. Large errors in the metadata can render them irrelevant to the network [case (i)].

Most work on community detection assumes that failure to find communities that correlate with metadata implies case (iv), algorithm failure, although some critical work has focused on case (iii), difficult or impossible to recover communities. The lack of consideration for cases (i) and (ii) suggests the possibility for selection bias in the published literature in this area (a point recently suggested by~\cite{hric2014community}). Indeed, recent critiques of the general utility of community detection in networks~\cite{leskovec2010empirical, yang2012community, hric2014community} can be viewed as a side effect of confusion about the role of metadata in evaluating algorithm results. 

\begin{figure}
	\centering
	\includegraphics[width=1.0\linewidth]{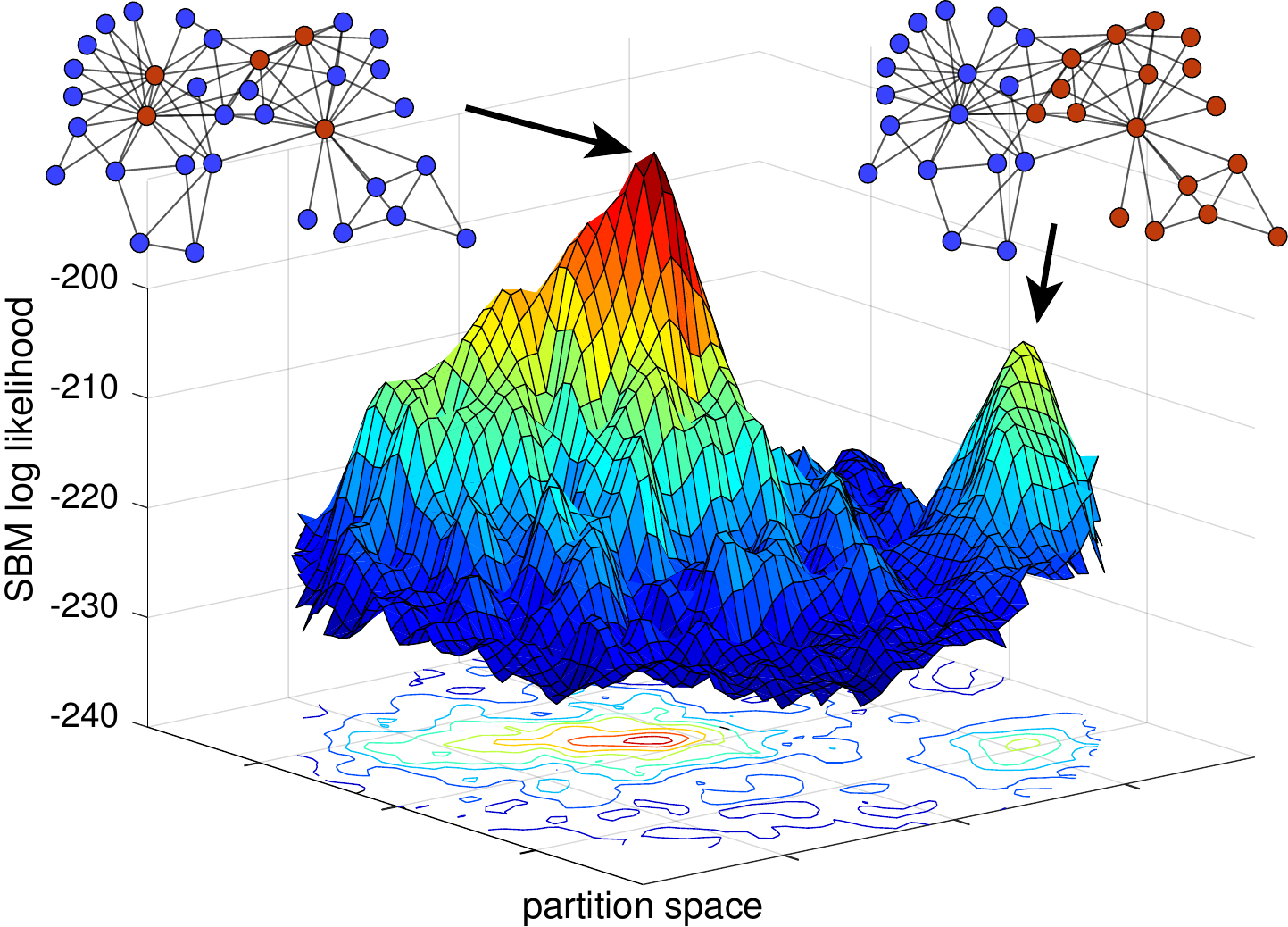}
	\caption{The stochastic blockmodel (SBM) log-likelihood surface for bipartitions of the Karate Club network~\cite{zachary1977information}. The high-dimensional space of all possible bipartitions of the network has been projected onto the $x,y$-plane (using a method described in Supplementary Text~\ref{sec_surfplots}) such that points representing similar partitions are closer together.  The surface shows two distinct peaks that represent scientifically reasonable partitions. The lower peak corresponds to the social group partition given by the metadata---often treated as ground truth---while the higher peak corresponds to a leader-follower partition.}
	\label{fig_karate_optima}
\end{figure}

For these reasons, using metadata to assess the performance of community detection algorithms can lead to errors of interpretation, false comparisons between methods, and oversights of alternative patterns and explanations, including those that do not correlate with the known metadata. 

For example, Zachary's Karate Club~\cite{zachary1977information} is a small real-world network with compelling metadata frequently used to demonstrate community detection algorithms. The network represents the observed social interactions of 34 members of a karate club. At the time of study, the club fell into a political dispute and split into two factions. 
These faction labels are the metadata commonly used as ground truth communities in evaluating community detection methods. However, it is worth noting at this point that Zachary's original network and metadata differ from those commonly used for community detection~\cite{girvan2002community}. Links in the original network were by the different types of social interaction that Zachary observed.  Zachary also recorded two metadata attributes: the political leaning of each of the members (strong, weak, or neutral support for one of the factions) and the faction they ultimately joined after the split. However, the community detection literature uses only the metadata representing the faction each node joined, often with one of the nodes mislabeled. This node (``Person number 9'') supported the president during the dispute but joined the instructor's faction as joining the president's faction would have involved retraining as a novice when he was only two weeks away from taking his black belt exam.

The division of the Karate Club nodes into factions is not the only scientifically reasonable way to partition the network. Figure~\ref{fig_karate_optima} shows the log-likelihood landscape for a large number of two-group partitions (embedded in two dimensions for visualization) of the Karate Club, under the stochastic blockmodel (SBM) for community detection~\cite{holland1983stochastic,nowicki2001estimation}.  Partitions that are similar to each other are embedded nearby in the horizontal coordinates, meaning that the two broad peaks in the landscape represent two distinct sets of high-likelihood partitions, one centered around the faction division and one that divides the network into leaders and followers. Other common approaches to community detection~\cite{girvan2002community,shi2000normalized}, suggest that the best divisions of this network have more than two communities~\cite{cheng2010uncovering,evans2010clique}. The multiplicity and diversity of good partitions illustrates the ambiguous status of the faction metadata as a desirable target. 
 
The Karate Club network is among many examples for which standard community detection methods return communities that either subdivide the metadata partition~\cite{krzakala2013spectral} or do not correlate with the metadata at all~\cite{karrer2011stochastic,newman2016structure}. 
More generally, most real-world networks have many good partitions and there are many plausible ways to sort all partitions to find good ones, sometimes leading to a large number of reasonable results. Moreover, there is no consensus on which method to use on which type of network~\cite{good2010performance,newman2016structure}. 

In what follows, we explore both the theoretical origins of these problems and the practical means to address the confounding cases described above. To do so, we make use of a generative model perspective of community detection. In this perspective, we describe the relationship between community assignements $\mathcal{C}$ and graphs $\mathcal{G}$ via a joint distribution $P(\mathcal{C},\mathcal{G})$ over all possible community assignments and graphs that we may observe.  We take this perspective because it provides a precise and interpretable description of the relationship between communities and network structure. Although generative models, like the SBM, describe the relationship between networks and communities directly via a mathematically explicit expression for $P(\mathcal{C},\mathcal{G})$, other methods for community detection nevertheless maintain an implicit relationship between network structure and community assignment. As such, the theorems we present, as well as their implications are more generally applicable across all methods of community detection.

In the next section we present rigorous theoretical results with direct implications for cases (i) and (iv), while the remaining sections introduce two statistical methods for addressing cases (i) and (ii), respectively. These contributions do not address case (iii), when there is no structure to be found, which has been previously explored by other authors, e.g., for the SBM~\cite{bordenave2015non, decelle2011prl, ghasemian2016detectability, massoulie2014community, mossel2014belief, mossel2015reconstruction} and modularity~\cite{guimera2004modularity, taylor2016detectability}.


\subsection*{Ground truth and metadata in community detection}
Community detection is an inverse problem: using only the edges of the network as data, we aim to find the grouping or partition of  the nodes that relates to how the network came to be. More formally, suppose that some data generating process $g$ embeds ground-truth communities $\mathcal{T}$ in the patterns of links in a network $\mathcal{G}=g(\mathcal{T})$. Our goal is to discover those communities based only on the observed links. To do so, we write down a community detection scheme $f$ that uses the network to find communities $\mathcal{C}=f(\mathcal{G})$. If we have chosen $f$ well, then the communities $\mathcal{C}$ will be equal to the ground truth $\mathcal{T}$ and we have solved the inverse problem. Thus, the community detection problem for a single network seeks a method $f^{*}$ that minimizes the distance between the identified communities and the ground truth:
\begin{equation}
	f^{*} = \argmin_f d(\mathcal{T},f(\mathcal{G}))\ ,
	\label{eq_minimizeT}
\end{equation}
where $d$ is a measure of distance between partitions. 

For a method $f$ to be generally useful, it should be the minimizer for many different graphs, each with its own generative process and ground truth. Often in the community detection literature, several algorithms are tested on a range of networks to identify which performs best overall~\cite{lancichinetti2009community,leskovec2010empirical, yang2015defining}. If a universally optimal community detection method exists, it must solve Eq.~\eqref{eq_minimizeT} for any type of generative process $g$ and partition $\mathcal{T}$, that is,
\begin{equation}
	\exists \ f^{*} \quad s.t. \quad f^{*} = \argmin_f d\big(\mathcal{T},f\left(g(\mathcal{T})\right)\big)\quad \forall \{g,\mathcal{T}\}\ .
	\label{eq_DL2}
\end{equation}
In fact, no such universal $f^{*}$ community detection method can exist because the mapping from generative models $g$ and ground truth partitions $\mathcal{T}$ to graphs $\mathcal{G}$ is not uniquely invertible due to the fact that the map is not a bijection. 
In other words, any particular network $\mathcal{G}$ can be produced by multiple, distinct generative processes, each with its own ground truth, such that $\mathcal{G} = g_{1}(\mathcal{T}_1) = g_{2}(\mathcal{T}_2)$, with $(g_1, \mathcal{T}_1) \neq (g_2,\mathcal{T}_2)$.  
Thus, no community detection algorithm method can uniquely solve the problem for all possible networks (Eq.~\eqref{eq_DL2}), or even a single network (Eq.~\eqref{eq_minimizeT}). This reasoning underpins the following theorem, which we state and prove in Supplementary Text~\ref{supp:bijection}:

\vspace{0.1in}
\begin{addmargin}[3em]{3em}
\noindent \textbf{Theorem 1}: For a fixed network $\mathcal{G}$, the solution to the ground truth community detection problem---given $\mathcal{G}$, find the $\mathcal{T}$ such that $\mathcal{G} = g(\mathcal{T})$---is not unique.
\end{addmargin}
\vspace{0.1in}

Substituting metadata $\mathcal{M}$ for ground truth $\mathcal{T}$ exacerbates the situation by creating additional problems. In real networks we do not know the ground truth or the generating process. Instead, it is common to seek a partition that matches some node metadata $\mathcal{M}$. Optimizing a community detection method to discover $\mathcal{M}$ is equivalent to finding $f^{*}$ such that
\begin{equation}
	f^{*} = \argmin_f \; d \! \left(\mathcal{M},f(\mathcal{G}) \right)\ ,
 	\label{eq_minimizeM} 
\end{equation}
yet this does not necessarily solve the community detection problem of Eq.~\eqref{eq_minimizeT} since we cannot guarantee that the metadata are equivalent to the unobserved ground truth, $d(\mathcal{M},\mathcal{T}) = 0$. Consequently, both $d(\mathcal{C},\mathcal{T})=0$ and $d(\mathcal{C},\mathcal{T})>0$ are possibilities. Thus, when we evaluate a community detection method by its ability to find a metadata partition, we confound the metadata's correspondence to the true communities, i.e., $d(\mathcal{M},\mathcal{T})$ [case (ii) in the previous section] and the community detection method's ability to find true communities, i.e., $d(\mathcal{C},\mathcal{T})$ [case (iv)]. In this way, treating metadata as ground truth simultaneously tests the metadata's relevance and the algorithm's performance, with no ability to differentiate between the two.
For instance, when considering competing partitions of the Karate Club (Figure~\ref{fig_karate_optima}), the leader-follower partition is the most likely partition under the SBM, yet it correlates poorly with the known metadata. On the other hand, under the degree-corrected SBM, the optimal partition is more highly correlated with the metadata (Fig.~\ref{fig_karate_neo}). Based only on the performance of recovering metadata, one would conclude that the degree-corrected model is better. However, if Zachary had not provided the faction information, but instead some metadata that correlated with the degree (e.g. the identities of the club's four officers) then our conclusion might change to the regular SBM being the better model. We would arrive at a different conclusion despite the fact that the network, and the underlying process that generated it, remain unchanged. A similar case of dependence on a particular choice of metadata are exemplified by divisions of high-school social networks using metadata of students' grade level or race \cite{newman2016structure}. Past evaluations of community detection algorithms that only measure performance by metadata recovery are thus inconclusive. It is only with synthetic data, where the generative process is known, that ground truth is knowable and performance objectively measurable.

However, even when the generative process is known for a single network or even a set of networks, there is no best community detection method over all networks. This is because, when averaged over all possible community detection problems, every algorithm has provably identical performance, a notion that is captured in a No Free Lunch theorem for community detection which we rigorously state and prove in Supplementary Text~\ref{supp:bijection} and paraphrase here:  

\vspace{0.1in}
\begin{addmargin}[3em]{3em}
\noindent \textbf{Theorem 3 (paraphrased)}: For the community detection problem, with accuracy measured by adjusted mutual information, the uniform average of the accuracy of any method $f$ over all possible community detection problems is a constant which is independent of $f$.
\end{addmargin}
\vspace{0.1in}

This No Free Lunch theorem, based on the No Free Lunch theorems for supervised learning~\cite{wolpert1996lack}, implies that no method has an \textit{a priori} advantage over any other across all possible community detection tasks. (In fact, Theorem 3 and its proof apply to clustering and partitioning methods in general, beyond community detection.) In other words, for a set of cases that a particular method $f_a$ outperforms $f_b$, there must exist a set of cases where $f_b$ outperforms $f_a$---on average no algorithm performs better than any other. Yet this does not render community detection pointless because the theorem also implies that if the tasks of interest correspond to a restricted subset of cases (e.g., finding communities in gene regulatory networks or certain kinds of groups in social networks), then there may indeed be a method that outperforms others within the confines of that subset. In short, matching beliefs about the data generating process $g$ with the assumptions of the algorithm $f$ can lead to better and more accurate results, at the cost of reduced generalizability. (See Supplementary Text~\ref{supp:bijection} for additional discussion.) 

The combined implications of the epistemological arguments in the previous section with Theorems 1 and 3 in this section do not render community detection impossible or useless, by any means. They do, however, imply that efforts to find a universally best community detection algorithm are in vain, and that metadata should not be used as a benchmark for evaluating or comparing the efficacy of community detection algorithms. These theorems indicate that better community detection results may stem from a better understanding of how to divide the problem space into categories of community detection tasks, eventually identifying classes of algorithms whose strengths are aligned with the requirements of a specific category.

\subsection*{Relating metadata and structure}

From a scientific perspective, metadata labels have direct and genuine value in helping to understand complex systems. Metadata describe the nodes, while communities describe how nodes interact. Therefore, correspondence between metadata and communities suggests a relationship between how nodes interact and the properties of the nodes themselves. This correspondence has been used productively to assist in the inference of community structure~\cite{newman2016structure}, learn the relationship between metadata and network topology~\cite{peel2011topological, peel2012supervised} and explain dependencies between metadata and network structure~\cite{fosdick2015testing}.

Here we propose two new methods to explore how metadata relate to the structure of the network when the metadata only correlate weakly with the identified communities. Both methods utilize the powerful tools of probabilistic models, but are not restricted to any particular model of community structure. The first method is a statistical test to assess whether or not the metadata partition and network structure are related [case (i)]. The second method explores the space of network partitions to determine if the metadata represent the same or different aspects of the network structure as the ``optimal'' communities inferred by a chosen model [case (ii)].

In principle, any probabilistic generative model (e.g.,~\cite{holland1983stochastic, nowicki2001estimation, airoldi2009mixed, ball2011efficient, larremore2014efficiently, peixoto2014hierarchical}) of communities in networks could be used within these methods. Here we derive results for the popular stochastic blockmodel~\cite{holland1983stochastic, nowicki2001estimation} and its degree-corrected version~\cite{karrer2011stochastic} (alternative formulations discussed in Supplementary Texts \ref{supp:BESTest} and \ref{supp:neoSBM}). The SBM defines communities as sets of nodes that are \textit{stochastically equivalent}. This means that the probability $p_{ij}$ of a link between a pair of nodes $i$ and $j$ depends only on their community assignment, i.e., $p_{ij} = \omega_{\pi_i,\pi_j}$, where $\pi_i$ is the community assignment for node $i$ and $\omega_{\pi_i,\pi_j}$ is the probability that a link exists between members of groups $\pi_i$ and $\pi_j$. This general definition of community structure is quite flexible, and allows for both assortative and disassortative community structure, as well as arbitrary mixtures thereof. 

\subsection*{Testing for a relationship between metadata and structure}\label{sec-BESTest}
Our first method, called the \textit{blockmodel entropy significance test} (BESTest), is a statistical test to determine if the metadata partition is relevant to the network structure [case (i)], i.e., if it provides a good description of the network under a given model. We quantify relevance using the entropy, which is a measure of how many bits of information it takes to record the network given both the network model and its parameters. The lower the entropy under this model, the better the metadata describe the network, while a higher entropy implies that the metadata and the patterns of edges in the network are relatively uncorrelated. We derive and discuss the BESTest using five different models in Supplementary Text~\ref{supp:BESTest}. Here we describe a particularly straightforward version of this test using the SBM. 

The BESTest works by first dividing a network's nodes according to the labels of the metadata and then computing the entropy of the SBM that best describes the partitioned nodes. This entropy is then compared to a distribution of entropies using the same network but random permutations of the metadata labels, resulting in a standard $p$-value. Specifically, we use the SBM with maximum likelihood parameters for the partition induced by the metadata, which is given by $\hat \omega_{rs} = \frac{m_{rs}}{n_{r}n_{s}}$ where $m_{rs}$ is the number of links between group $r$ and group $s$ and $n_r$ is the number of nodes in group $r$. Then we compute the entropy $H(\mathcal{G};\mathcal{M})$, which we derive and discuss in detail, along with derivations of entropies for other models, in Supplementary Text~\ref{supp:BESTest}.

The statistical significance of the entropy value $H(\mathcal{G};\mathcal{M})$ is obtained by comparing it to the entropy of the same network but randomly permuted metadata. Specifically, we compute a null distribution of such values, derived by calculating the entropies induced by random permutations $\{\tilde{\pi}\}$ of the observed metadata values $H(\mathcal{G};\tilde{\pi})$. This choice of null model preserves both the empirical network structure and the relative frequencies of metadata values, but removes the correlation between the two. The result is a standard $p$-value, defined as 
\begin{equation}
	p\textrm{-value} = \text{Pr}\left [ H(\mathcal{G};\tilde{\pi}) \leq H(\mathcal{G};\mathcal{M}) \right ]\,,
	\label{eq-Q}
\end{equation} 
which can be estimated empirically by computing $H(\mathcal{G};\tilde{\pi})$ for a large number of randomly permuted metadata vectors $\tilde \pi$. Smaller $p$-values indicate that the metadata provide a better description of the network, making it relatively less plausible that a random permutation of the metadata values could describe the network as well as the observed metadata does.  It is important to note that $p$-values measure statistical significance but not effect strength, meaning that a low $p$-value does not indicate a strong correlation between the metadata and the network structure. Recently, Bianconi \textit{et al}.~\cite{bianconi2009assessing} proposed a related entropy test for this task, based on a Normal approximation to the null distribution under the SBM. The blockmodel entropy significance test described here is a generalization of Bianconi et al.'s test that is both more flexible, as it can be used with any number of null models, and more accurate, as the true null distribution is substantially non-Normal (Fig.~\ref{fig-BESTest}).

\begin{figure}[t]
\centering
	\includegraphics[width=\linewidth]{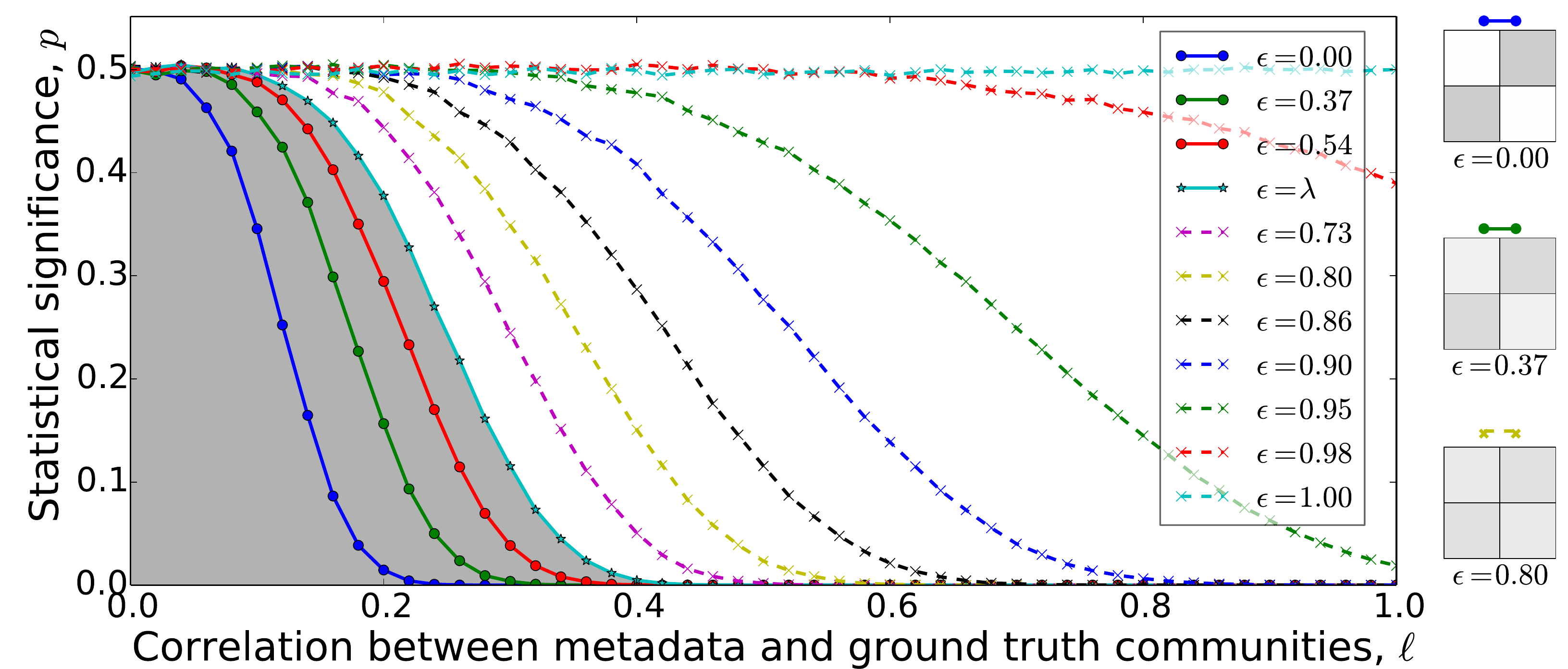}
	\caption{Expected $p$-value estimates of the blockmodel entropy significance test as the correlation  $\ell$ between metadata and planted communities increases (each metadata label correctly reflects the planted community with probability $(1+\ell)/2$; see Supplementary Text \ref{supp:BESTest}). Each curve represents networks with a fixed community strength $\epsilon=\omega_{rs} / \omega_{rr}$. Solid lines indicate strong community structure in the so-called detectable regime ($\epsilon<\lambda$), while dashed lines represent weak undetectable communities ($\epsilon>\lambda$)~\cite{decelle2011prl}. Three block density diagrams visually depict $\epsilon$ values.} 
	\label{fig-BESTest_sensitivity}
\end{figure}

The blockmodel entropy significance test is, by construction, sensitive to even low correlations between metadata and network structure. To quantify the sensitivity of this $p$-value, we first apply it to synthetic networks with known community structure (see Supplementary Text \ref{supp:BESTest} for a complete description of synthetic network generation). For these networks, our ability to detect relevant metadata is determined jointly by the strength of the planted communities and the correlation between metadata and communities. Figure~\ref{fig-BESTest_sensitivity} shows that for networks with strong community structure we can reliably detect relevant metadata even for relatively low levels of correlation with the planted structure. In fact, our method can still identify relevant metadata when the community structure is sufficiently weak that communities are provably undetectable by any community detection algorithm that relies only on the network~\cite{decelle2011prl}. Statistical significance requires an increasing level of correlation with the underlying structure as community strength decreases; if there is no structure in the network ($\epsilon=1$) then any metadata partition will be correctly identified as irrelevant. Note that a low $p$-value does not mean that the metadata provide the best description of the network, nor does it imply that we should be able to recover the metadata partition using community detection.

We now apply the blockmodel entropy significance test to a social network of interactions within a law firm, and to biological networks representing similarities among genes in the human malaria parasite \textit{P. falciparum} (see Supplementary Text \ref{supp:data}). The first set, the Lazega Lawyers networks, comprises three networks on the same set of nodes and five metadata attributes. The multiple combinations of edge and metadata types that yield highly significant $p$-values (Table~\ref{tab:pvals_main_lazega}; see Table~\ref{tab:pvals} for results using additional models of community structure) indicate that each set of metadata provides non-trivial information about the structure of multiple networks, and vice versa, implying that all metadata sets are relevant to the edge formation process, so none should be individually treated as ground truth.

\begin{table}[t]
 \caption{BESTest $p$-values for Lazega Lawyers}
 \label{tab:pvals_main_lazega}
 \begin{center}
 \begin{tabular}{l|ccccccccc} 
 \hline
 & \multicolumn{5}{c}{Metadata Attribute} \\ 
 Network & Status & Gender & Office & Practice & Law School \\
 \hline
Friendship	&$<10^{-6}$	&$0.034$	&$<10^{-6}$	&$0.033$	&$0.134$	\\
Cowork	&$<10^{-3}$	&$0.094$	&$<10^{-6}$	&$<10^{-6}$	&$0.922$	\\
Advice	&$<10^{-6}$	&$0.010$	&$<10^{-6}$	&$<10^{-6}$	&$0.205$	\\
 \hline
 \end{tabular}

\vspace{0.1in}

\caption{BESTest $p$-values for Malaria {\it var} genes}
 \label{tab:pvals_main_malaria}
\begin{tabular}{l|ccccccccc} 
 \hline
&\multicolumn{9}{c}{{\it var} Gene Network Number} \\ 
& 1 & 2 & 3 & 4 & 5 & 6 & 7 & 8 & 9 \\
 \hline
Genome
&$0.566$
&$0.064$
&$0.536$
&$0.588$
&$0.382$
&$0.275$
&$0.020$
&$0.464$
&$0.115$\\
 \hline
 \end{tabular}
 \end{center}
\end{table}

The second set, the malaria \textit{var} gene networks, comprises nine networks on the same set of nodes and three sets of metadata. For each network, we find a non-significant $p$-value when the metadata denote the parasite genome-of-origin (Table~\ref{tab:pvals_main_malaria}; see Table~\ref{tab:mal_pvals} for results using additional models of community structure and additional metadata). In contrast to the Lazega Lawyers network, these genome metadata are statistically irrelevant for explaining the observed patterns of gene recombinations. This finding substantially strengthens the conclusions of Ref.~\cite{larremore2013network} which used a less sensitive test based on label assortativity. Some metadata for these networks do correlate, however (see Supplementary Text~\ref{supp:BESTest}).
\begin{figure*}[t]
 \centering
 \includegraphics[width=1.0\textwidth]{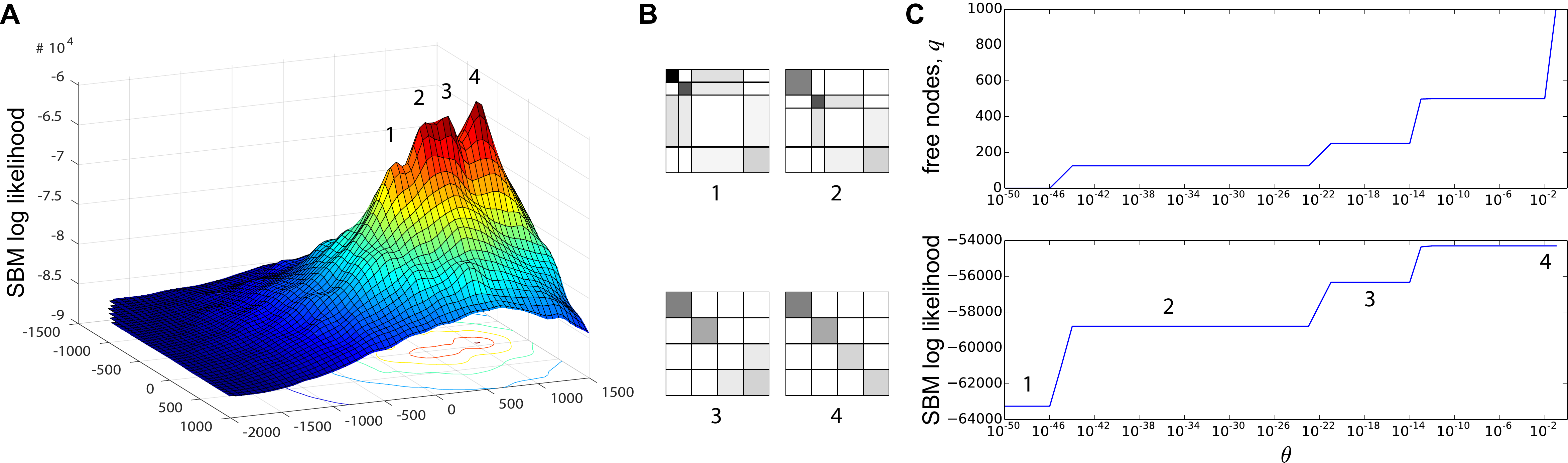} 
 \caption{The neoSBM on synthetic data. (A) The stochastic block model (SBM) likelihood surface shows four distinct peaks corresponding to a sequence of locally optimal partitions. (B) Block density diagrams depict community structure for locally optimal partitions, where darker color indicates higher probability of interaction. (C) The neoSBM, with partition 1 as the metadata partition, interpolates between partition 1 and the globally optimal SBM partition 4. The number of free nodes $q$ and SBM log likelihood as a function of $\theta$ show three discontinuous jumps as the neoSBM traverses each of the locally optimal partitions (1--4).}
 \label{fig_synth_neo}
\end{figure*}

\subsection*{Diagnosing the structural aspects captured by metadata and communities}
\label{sec_neoSBM}
Our second method provides a direct means to diagnose whether some metadata and a network's detected communities differ because they reveal different aspects of the network's structure [case (ii)]. We accomplish this by extending the SBM to probe the local structure around and between the metadata partition and the detected structural communities. This extended model, which we call the {\it neoSBM}, performs community detection under a constraint in which each node is assigned one of two states, which we call blue or red, and a parameter $q$ that governs the number of nodes in each state. If a node is blue, its community is fixed as its metadata label, while if it is red, its community is free to be chosen by the model. We choose $q$ automatically within the inference step of the model by imposing a likelihood penalty in the form of a Bernoulli prior with parameter $\theta$, which controls for the additional freedom that comes from varying $q$. The neoSBM's log likelihood is $\mathcal{L}_{\textrm{neoSBM}} = \mathcal{L}_{\textrm{SBM}} + q\psi(\theta)$, where $\psi(\theta)$ may be interpreted as the cost of freeing a node from its metadata label (see Supplementary Text \ref{supp:neoSBM} for exact formulation).

By varying the cost of freeing a node, we can use the neoSBM to produce a graphical diagnostic of the interior of the space between the metadata partition and the inferred community partition. In this way, the neoSBM can shed light on how the metadata and inferred community partitions are related, beyond direct comparison of the partitions via standard techniques such as normalized mutual information or the Rand index. As the cost of freeing nodes is reduced, the neoSBM creates a path through the space of partitions from metadata to the optimal community partition and, as it does so, we monitor the improvement of the partition by the increase in SBM log likelihood. A steady increase indicates that the neoSBM is incrementally refining the metadata partition until it matches the globally optimal SBM communities. This behavior implies that the metadata and community partitions represent related aspects of the network structure. On the other hand, if the SBM likelihood remains constant for a substantial range of $\theta$, followed by a sharp increase or jump, it indicates that the neoSBM has moved from one local optimum to another. Multiple plateaus and jumps indicate that several local optima have been traversed, revealing that the partitions are capturing different aspects of the network's structure.

To demonstrate the usage of the neoSBM, we examine the path between partitions for a synthetic network with four locally optimal partitions which correspond to the four distinct peaks in the surface plot (Fig.~\ref{fig_synth_neo}A; see Supplementary Text \ref{supp:neoSBM} for a complete description of synthetic network generation). We take the partition of the lowest of these peaks as metadata and use the neoSBM to generate a path to the globally optimal partition by varying the $\theta$ parameter of the neoSBM from $0$ to $1$. The corresponding changes in the SBM log likelihood and the number of free nodes show three discontinuous jumps (Fig.~\ref{fig_synth_neo}C), one for each time the model encounters a new locally optimal partition.

Examining the partitions along the neoSBM's path can provide direct insights into the relationship between metadata and network structure. Figure~\ref{fig_synth_neo}B shows the structure at each of the four traversed optima as block-wise interaction matrices $\omega$. Each partition has a different type of large-scale structure, from core-periphery to assortative patterns. In this way, when metadata do not closely match inferred communities, the neoSBM can shed light on whether and how the metadata capture similar or different aspects of network structure.

\begin{figure*}[t]
\centering
 \includegraphics[width=1.0\textwidth]{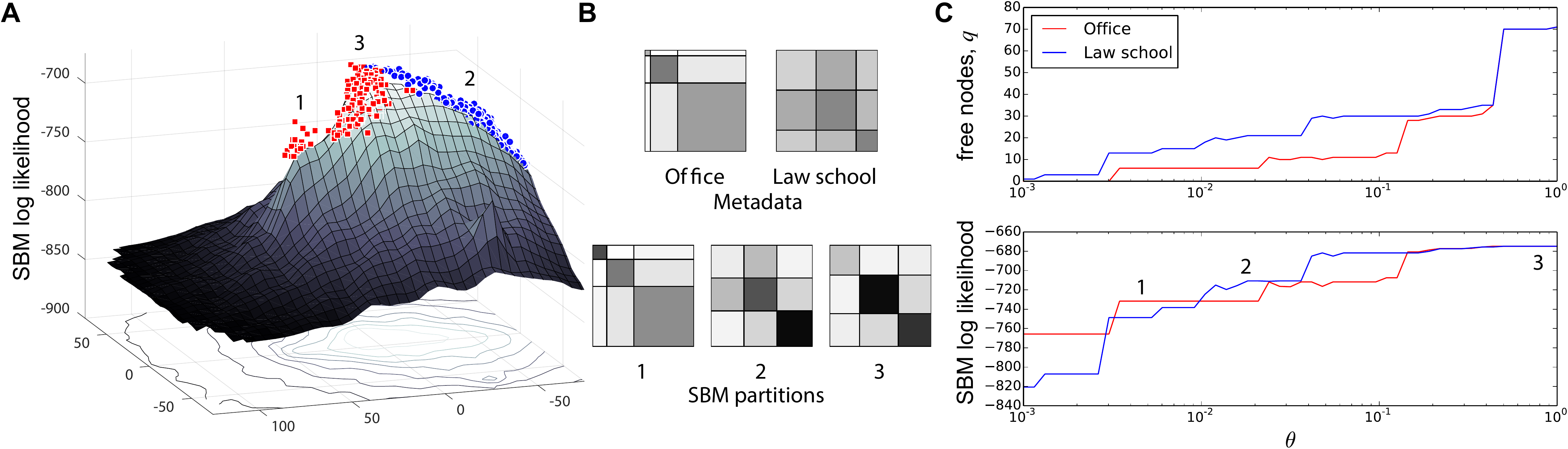}
\caption{The neoSBM on Lazega Lawyers friendship data~\cite{lazega2001collegial}. (A) Points of two neoSBM paths using office (red) and law school (blue) metadata partitions are shown on the stochastic block model (SBM) likelihood surface (greyscale to emphasize paths). (B) Block density diagrams depict community structure for metadata, (1--2) intermediate optimal, and (3) globally optimal partitions, where darker color indicates higher probability of interaction. (C) The neoSBM traverses two distinct paths to the global optimum (3), but only the path beginning at the office metadata partition traverses a local optimum (1), indicated by a plateau in free nodes $q$ and log likelihood.}
 \label{fig_laz_neo}
\end{figure*}

We now present an application of the neoSBM to the Lazega Lawyers data analyzed in the previous section. When initialized with the law school and office location metadata, the neoSBM produces distinct patterns of relaxation to the global optimum (Fig.~\ref{fig_laz_neo}A,C), approaching it from opposite sides of the peak in the likelihood surface. Starting at the law school metadata, the model traverses the space of partitions to the global SBM-optimal partition without encountering any local optima. In contrast, the path from the office metadata crosses one local optimum (Fig.~\ref{fig_laz_neo}A,B), which indicates that the law school metadata are more closely associated with the large-scale organization of the network than are the office metadata. Both sets of metadata labels are relevant, however, as we determined in the previous section using the BESTest. Results for other real-world networks are included in Supplementary Text \ref{supp:neoSBM}, including generalizations of the neoSBM to other community detection methods.

\section*{Discussion}
Treating node metadata as ground truth communities for real-world networks is commonly justified via an erroneous belief that the purpose of community detection is to recover groups that match metadata labels~\cite{ahn2010link,yang2012community, yang2015defining, hric2014community}. Consequently, metadata recovery is often used to measure community detection performance~\cite{soundarajan2012using} and metadata are often referred to as ground truth~\cite{chakraborty2013computer,newman2016structure}. However, the organization of real networks typically correlates with multiple sets of metadata, both observed and unobserved. Thus, labeling any particular set to be ``ground truth'' is an arbitrary and generally unjustified decision. Furthermore, when a community detection algorithm fails to identify communities that match known metadata, poor algorithm performance is indistinguishable from three alternative possibilities: (i) the metadata are irrelevant to the network structure, (ii) the metadata and communities capture different aspects of the network structure, or (iii) the network lacks group structure. Here, we have introduced two new statistical tools to directly investigate cases (i) and  (ii), while (iii) remains well addressed by work from other authors~\cite{bordenave2015non, decelle2011prl, ghasemian2016detectability, massoulie2014community, mossel2014belief, mossel2015reconstruction,guimera2004modularity, taylor2016detectability}. We have also articulated multiple mathematical arguments which conclude that treating metadata as ground truth in community detection induces both theoretical and practical problems. However, we have also shown that metadata remain useful and that a careful exploration of the relationship between node metadata and community structure can yield new insights into the network's underlying generating process.

By searching only for communities that are highly correlated with metadata, we risk focusing only on positive correlations while overlooking other scientifically relevant organizational patterns. In some cases, disagreements between metadata labels and community detection results may in fact point to interesting or unexpected generative processes. For instance, in the Karate Club network, there is one node whose metadata label is not recovered by most algorithms. A close reading of Zachary's original manuscript reveals that there is a rational explanation for this one-node difference: although the student had more social ties to the president's group, he chose to join the instructor's group so as not to lose his progress toward his black belt~\cite{zachary1977information}. In other cases, metadata may provide a narrative that blinds us to additional structure, exemplified by a network of political blogs~\cite{adamic2005political} in which liberal and conservative blogs formed two highly assortative groups. Consequently, recovery of these two groups has been used as a signal that a method produces ``good'' results~\cite{karrer2011stochastic}. A deeper analysis, however, suggests that this network is better described by subdividing these two groups, a step that reveals substantial substructure within the dominant patterns of political connectivity~\cite{krzakala2013spectral, peixoto2014hierarchical}. These subgroups remained overlooked in part because the metadata labels aligned closely with an attractively simple narrative. 

The task of community detection is the network analog of data clustering. Whereas clustering divides a set of vectors into groups with similar attribute patterns, community detection divides a network into groups of nodes with similar connectivity patterns. The general problem of clustering, however, is notoriously slippery~\cite{vonluxburg2012clustering} and cannot be solved universally~\cite{kleinberg2003impossibility}. Essentially, which clustering is optimal depends on its subsequent uses, and our theoretical results here show that similar constraints hold for community detection~\cite{browet2016incompatibility}. However, as with clustering, despite the lack of a universal solution, community detection remains a useful and powerful tool in the analysis of complex networks.

There is no universally accepted definition of community structure, nor should there be. Networks represent a wide variety of complex systems, from biological to social to artificial systems, and their large-scale structure may be generated by fundamentally different processes. Good community detection methods like the SBM can be powerful exploratory tools, able to uncover a wide variety of such patterns in real networks. But, as we have shown here, there is no free lunch in community detection. Instead, algorithmic biases that improve performance on one class of networks must reduce performance on others. This is a natural trade off, and suggests that good community detection algorithms come in two flavors: general algorithms that perform fairly well on a wide variety of tasks and inputs, and specialized algorithms that perform very well on a more narrow set of tasks, outperforming any general algorithm, but which perform more poorly when applied outside their preferred domain [an insight foreshadowed in past work~\cite{clauset2008hierarchical}]. In fact, in some cases it may be advantageous to use a set of carefully chosen metadata and a narrow set of corresponding networks to train specialized algorithms. Historically, most work on community detection algorithms has focused on developing general approaches. A deeper consideration of how the outputs of community detection algorithms will be subsequently used, e.g., in testing scientific hypotheses, predicting missing information, or simply coarse-graining the network, may shed new light on how to design better algorithms for those specific tasks. An important direction of future work is thus to better understand both these trade offs and the errors that can occur in domain-agnostic applications~\cite{peel2011estimating, yang2016comparative}. 

A complementary approach is to incorporate the metadata into inference process itself, which can help guide a method toward producing more useful results.  The neoSBM introduced here is one such method. Others include methods that use metadata as a prior for community assignment~\cite{newman2016structure} and identify relevant communities to  predict missing network or metadata information~\cite{peel2011topological, peel2012supervised, hric2016network}. However, there is potential to go further than these domain-agnostic methods can take us. Tools that incorporate correct domain-specific knowledge about the systems they represent will provide the best lens for revealing patterns beyond what is already known and ultimately lead to important scientific breakthroughs. By rigorously probing these relationships we can move past the false notion of metadata as ground truth, and instead uncover the particular organizing principles underlying real world networks and their metadata.

\section*{Acknowledgements}
The authors thank Cris Moore, Tiago Peixoto, Michael Schaub, David Wolpert, and Johan Ugander for insightful conversations. This work was supported by IAP (Belgian Scientific Policy Office) and ARC (Federation Wallonia-Brussels) (LP), the SFI Omidyar Fellowship (DBL), and NSF Grant IIS-1452718 (AC).

\section*{Data and code} Computer code implementing the analysis methods described in this paper and other information can be found online at \href{https://piratepeel.github.io/code.html}{https://piratepeel.github.io/code.html} and \href{http://danlarremore.com/metadata}{http://danlarremore.com/metadata}.

\bibliography{bibliography_metadata}

\appendix
\renewcommand{\thefigure}{S\arabic{figure}}
\renewcommand{\thetable}{S\arabic{table}}
\setcounter{figure}{0}
\setcounter{table}{0}

\clearpage
\section{The neoSBM}\label{supp:neoSBM}

\noindent\matrixquote{{\bf Morpheus}: \textit{Unfortunately, no one can be told what the Matrix is. You have to see it for yourself... This is your last chance. After this, there is no turning back. You take the blue pill, the story ends, you wake up in your bed and believe whatever you want to believe. You take the red pill, you stay in Wonderland, and I show you how deep the rabbit hole goes. Remember: all I'm offering is the truth. Nothing more.}~\cite{matrix1999}
}

This Supplementary Text is divided into four subsections providing additional details on the neoSBM. 
\begin{itemize}
\item Subsection I describes the neoSBM (I.a) and the inference methods used in this paper (I.b). 
\item Subsection II describes the generation of the synthetic network used in the main text, Fig.~\ref{fig_synth_neo}. 
\item Subsection III describes how the neoSBM can be extended to other models including the degree corrected neoSBM. 
\item Subsection IV provides additional examples of results of the neoSBM applied to the Lazega Lawyers networks (IV.a) and the Malaria networks (IV.b).
\end{itemize}

For convenience, we provide a reference table of notation used in derivations in this Supplementary Text.

\begin{table}[h!]
 \caption{Notation used in this Supplementary Text}
 \scriptsize
 \begin{center}
 \begin{tabular}{c|l} 
 \hline
Variable & Definition \\
 \hline
$\mathcal{G}$ & a network, $\mathcal{G}=(V,E)$  \\
$N$ & the number of nodes $\vert V \vert$\\
$A_{ij}$ & the number of edges between nodes $i$ and $j$, $A_{ij} \in \{0,1\} $\\
$k_i$ & the degree of node $i$. \\
$\omega_{rs}$ & the probability of an edge between nodes in groups $r$ and $s$ \\
$\pi$ & a partition of nodes into groups \\
$M$ & a set of metadata labels \\
$C$ & an inferred optimal community assignment \\
$z$ & neo-state indicator variable, $z_i \in \{b,r\}$ \\
$\theta$ & Bernoulli prior probability parameter \\
$\mathcal{L}_{X}$ & log likelihood $L$ of model $X$ \\
$q$ & the number of free nodes, $q=\sum_i \delta_{z_i,r}$ \\
$\delta_{a,b}$ & the Kronecker delta: $\delta_{a,b}=1$ for $a = b$; $\delta_{a,b}=0$ for $a \neq b$ \\
 \hline
 \end{tabular}
 \end{center}
\end{table}

\subsection{neoSBM model description and inference}

\subsubsection{Model description}

The neoSBM extends the SBM, allowing metadata to influence the inferred partitions by controlling the number of nodes that are assigned to groups according to their metadata labels. The task of the neoSBM is to perform community detection under a constraint in which each node is assigned a latent state variable $z_i$, which can take one of two states, which we call blue or red. If a node is blue $z_i=b$, its community is fixed as its metadata label $\pi_i=M_i$.  However, if it is red $z_i=r$, its community is free to be chosen by the model. 
 We adjust the number of free nodes $q$ by varying the Bernoulli prior probability $\theta$ that a node will be free (red state). We can then write down the likelihood $L_{\rm{neo}}$ of a network $\mathcal{G}$ given a community assignment $\pi$ under the neoSBM as:
\begin{equation}
 L_{\rm{neo}}(\mathcal{G};\pi, z)=\prod_{ij}{\omega_{\pi_i\pi_j}^{A_{ij}}(1-\omega_{\pi_i\pi_j})^{(1-A_{ij})}}
 \prod_{i}{\theta^{\delta_{z_i,r}}(1-\theta)^{\delta_{z_i,b}}} \ .
 \label{eq_neoSBM-lik}
\end{equation} 
The first product in Eq.~\eqref{eq_neoSBM-lik} corresponds to the standard SBM likelihood $L_{\rm{sbm}}$, while the second product corresponds to the probability of the states $P(z=r|\theta)$ and acts as a penalty function to control the number of free nodes. While it is possible to find communities by optimizing Eq.~\eqref{eq_neoSBM-lik} directly, instead we work with the more practical log likelihood,
\begin{align}
 \mathcal{L}_{\rm{neo}}(\mathcal{G};\pi, z)= & \sum_{ij}{A_{ij} \log \omega_{\pi_i\pi_j} + (1-A_{ij}) \log (1-\omega_{\pi_i\pi_j})}  \notag \\ 
 & + \sum_{i}{\delta_{z_i,r} \log \theta + \delta_{z_i,b} \log (1-\theta)} \enspace,
 \label{eq_neoSBM-loglik}
\end{align} 
since maximizing Eq.~\eqref{eq_neoSBM-lik} is equivalent to maximizing Eq.~\eqref{eq_neoSBM-loglik}. We can then rearrange the second sum $\log P(z=r|\theta)$, to give:
\begin{align}
 \log P(z=r|\theta) = & \; \sum_i \delta_{z_i,r} \left(\log \frac{\theta}{1-\theta} \right) + N \log(1-\theta) \notag \\
 = & \; q \psi(\theta) + N \log(1-\theta) \enspace ,
\end{align}
dropping the constant term, we can rewrite the neoSBM log likelihood in terms of the SBM log likelihood and a function of the number of free nodes $q$,
\begin{equation}
	\mathcal{L}_{\rm{neo}}(\mathcal{G};\pi, z) = \mathcal{L}_{\rm{sbm}}(\mathcal{G};\pi) + q \psi(\theta) \enspace .
\end{equation}

We emphasize that in the equation above, $\theta$ is a fixed parameter, and $q$ is selected automatically during inference as part of the likelihood maximization. Optimization of $\mathcal{L}_{\rm{sbm}}$ yields the SBM optimal communities $C$,
\begin{equation}
	C= \arg \max_{\pi} \mathcal{L}_{\rm{sbm}}(\mathcal{G};\pi) \enspace ,
\end{equation}
and so the SBM likelihood given the metadata partition $M$ will always be less than or equal to the likelihood of the inferred partition $C$.  That is $\mathcal{L}_{\rm{sbm}}(\mathcal{G};M) \leq \mathcal{L}_{\rm{sbm}}(\mathcal{G};C)$, where the inequality is saturated if and only if the metadata is equal to the optimal SBM partition. So the minimum number of free nodes $\hat{q}$ required to maximize the SBM likelihood is
\begin{equation}
 \hat{q} = \sum_{i}{1 - \delta_{M_i,C_i}} \enspace ,
\end{equation} 
for which the label permutations of $M$ and $C$ are maximally aligned.  Whenever $q> \hat{q}$ there will be no further improvement in $\mathcal{L}_{\rm{sbm}}$.  To interpolate between $M$ and $C$ we vary the prior probability of each node to take the red state $P(z=r|\theta)$.  For values of $\theta<0.5$ we can interpret the log probability, or $\psi(\theta)$, as the cost of freeing a node because the log likelihood $\mathcal{L}_{\rm{neo}}$ will incur a penalty for setting each $z_i=r$.   Maximizing $\mathcal{L}_{\rm{neo}}$ is therefore a trade-off between freeing nodes to maximize $\mathcal{L}_{\rm{sbm}}$ and fixing nodes to metadata labels to maximize $\log P(z|\theta)$. 
 When the SBM likelihood of both partitions is equal (i.e., $M=C$) then $\mathcal{L}_{\text{neo}}(\mathcal{G};\pi, z)$ will be maximized when $q=0$ unless $\theta \geq 0.5$.  However, when $\mathcal{L}_{\text{sbm}}(\mathcal{G};M) < \mathcal{L}_{\text{sbm}}(\mathcal{G};C)$, $q$ can be greater than $0$ if the resulting partition $\pi$ provides a sufficient increase in log likelihood.  Specifically, if 
\begin{equation}
 \mathcal{L}_{\text{sbm}}(\mathcal{G};\pi) - \mathcal{L}_{\text{sbm}}(\mathcal{G};M) 
> q \psi(\theta) \enspace , 
\end{equation}
then it indicates that the cost of freeing $q$ nodes is outweighed by its contribution to improving the likelihood.

Here we have discussed the extension of the SBM to the neoSBM, but this extension can be easily generalized to any probabilistic generative network model that specifies the likelihood of a graph given a partition of the network. We present one such generalization, the degree-corrected neoSBM, in subsection III of this Supplementary Text.

\subsubsection{Inference}
Inference of the parameters of the neoSBM was performed using a Markov chain Monte Carlo (MCMC) approach. The community labels of the free nodes were inferred in the same way as the standard SBM~\cite{peixoto2014efficient}. However, to infer the values of $z_i$ that determined whether or not each node was free, we used a uniform Bernoulli (i.e., a fair coin) as a proposal distribution. Since this distribution is symmetric we  can simply accept each proposal with probability $a$:
\begin{equation}
	 a = \min \left\{\Delta L_{\rm{neo}},1 \right\} \enspace.
\end{equation} 

To avoid getting trapped in local optima of the likelihood, we initialize the neoSBM with the labels set to the inferred SBM partition, $\pi=C$, and all nodes initialized to be free, $z_i=r$ for all $i$.

\subsection{Extensions}
The neoSBM can easily be extended to any probabilistic model for which we identify communities by maximizing the model likelihood.  As an example, consider the degree-corrected SBM, which allows for nodes with heterogenous degrees to belong to the same community (see Supplementary Text \ref{supp:BESTest} for more details). We can create a degree-corrected neoSBM in much the same way as we created the neoSBM, by penalizing the likelihood according to the number of free nodes using a Bernoulli prior.  This treatment gives the log likelihood:
\begin{equation}
	\mathcal{L}_{\rm{dcneo}}(\mathcal{G};\pi, z) = \mathcal{L}_{\rm{dcsbm}}(\mathcal{G};\pi) + q \psi(\theta) \enspace ,
\end{equation}
 where $q \psi(\theta) = q \log P(z=r|\theta) + N \log(1-\theta)$ as before.  We present results from this model in subsection IV of this Supplementary Text. 
 
 We can also easily extend the neoSBM to other, non-probabilistic, community detection methods provided they explicitly optimize a global objective function.  Then we can similarly create a penalized version of this objective function.  That is, for some community detection model $X$, we can create a \textit{neo}-objective function $\mathcal{U}_{neoX}$
\begin{equation}
 \mathcal{U}_{neoX} = \mathcal{U}_{X} + q \psi(\theta) \enspace ,	
\end{equation}
where $\psi(\theta)$ could either represent the Bernoulli prior as before or any other cost function, e.g., $\psi(\theta)=\theta$, for $\theta \leq 0$.

\subsection{IV. Results on real-world networks}

\begin{figure*}[ht]
	\centering
	\includegraphics[width=\textwidth]{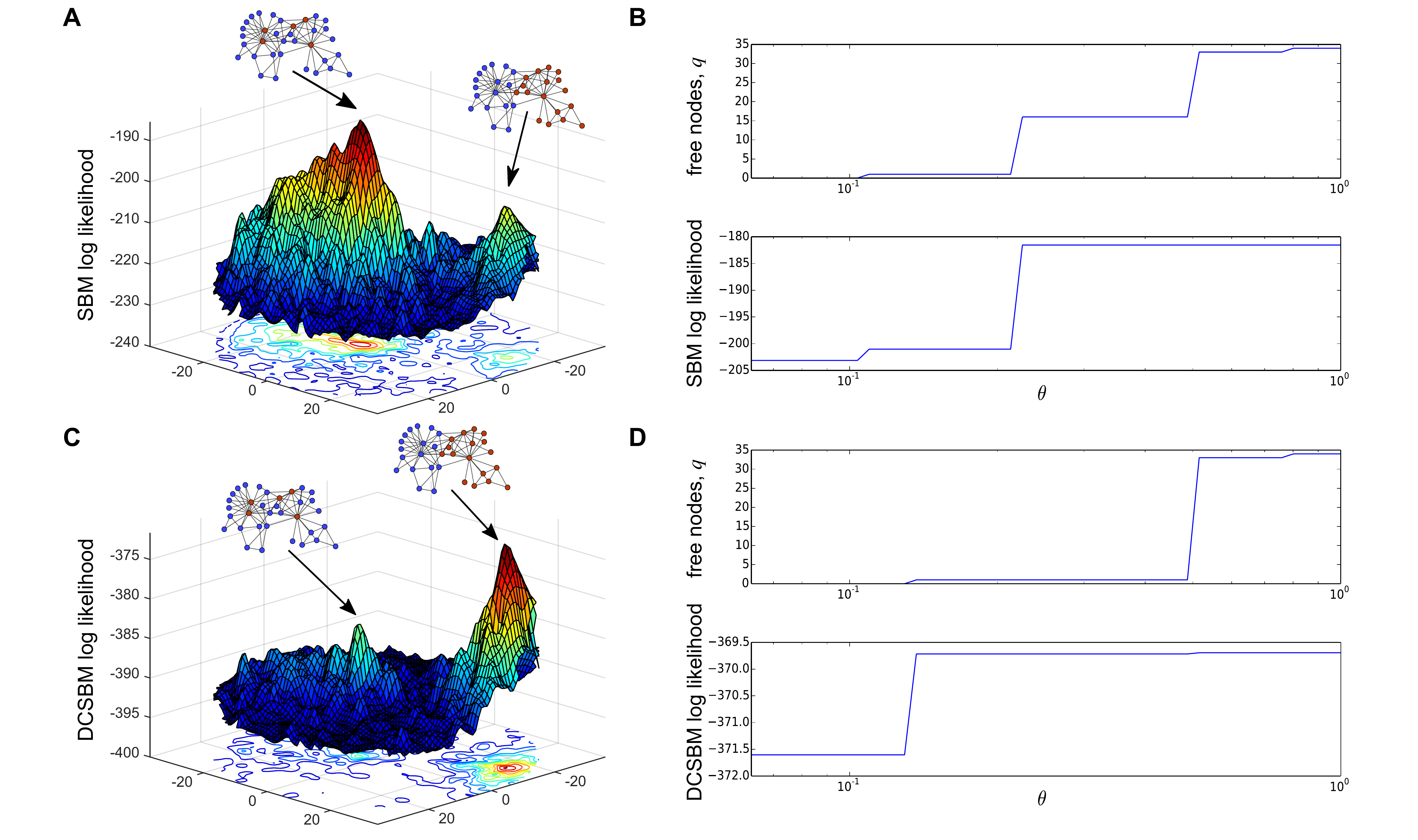}
	\caption{The results of the neoSBM and the degree-corrected neoSBM on the karate club network.  The SBM and DCSBM log likelihood surfaces (A and C respectively) show distinct two peaks that correspond to the same two partitions of the network: the two social factions and the leader-follower partition.  When we use the faction partition as metadata, we  from the output (B and D) that both models change a single node in order to reach the locally optimal partition.  For the neoDCSBM (D), this is the global optimum and no further change is observed.  For the neoSBM, the leader-follower partition is globally optimal, so once theta is large enough we see the model jump to this partition.}
	\label{fig_karate_neo}
\end{figure*}

In order to further demonstrate the neoSBM and the neoDCSBM described above, we present and discuss the application of the neoSBM to malaria {\it var} gene networks  and the application of the neoDCSBM to the Karate Club network. Full details about these data sets are presented in Supplementary Text \ref{supp:data}.

\subsubsection{neoDCSBM and the Karate Club network}
The likelihood surface for both models contains two local optima that correspond two the same two partitions, each being globally optimal for one of the models. Using the faction each member joined after the club split as metadata Fig.~\ref{fig_karate_neo} compares the output from the neoSBM and the neoDCSBM. Both models initially change just a single node to reach a local optimum. For the DCSBM this is the global optimum and so we see no further change. However, for the neoSBM this is not the global optimum (see Fig.~\ref{fig_karate_optima}) and so once $\theta$ is large enough we see a discontinuous jump as it switches to the globally optimal high-degree/low-degree partition.

\subsubsection{neoSBM and the Malaria \textit{var} gene networks}
\begin{figure*}[ht]
	\centering
	\includegraphics[width=\textwidth]{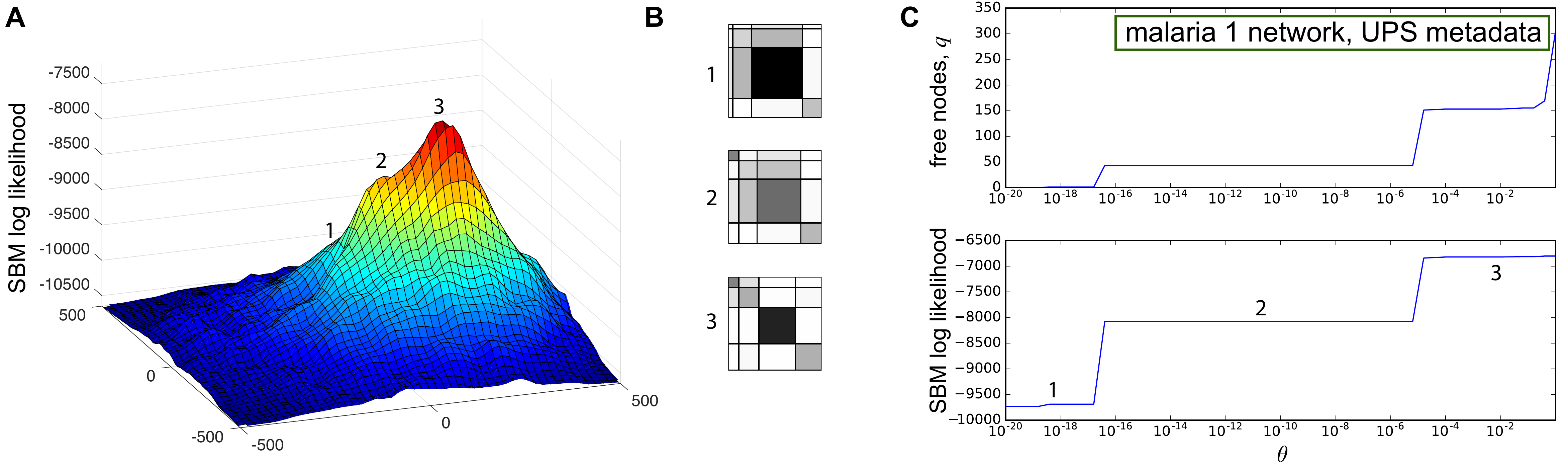}
	\caption{Results of the neoSBM on the malaria {\it var} gene network at locus one (``malaria 1'') using UPS metadata. (A) The SBM likelihood surface shows two peaks, one subtle 2 and one prominent 3, corresponding to a locally optimal partition near the metadata and the globally optimal partition, respectively. There is no peak at the metadata partition 1, however.  (B) Block density diagrams depict community structure for metadata and locally optimal partitions, where darker color indicates higher probability of interaction. (C) The neoSBM, beginning from UPS metadata, interpolates between metadata 1 and the globally optimal SBM partition 3. The number of free nodes $q$ and SBM log likelihood as a function of $\theta$ shows two discontinuous jumps as the neoSBM traverses from the metadata to the locally optimal partition ($1 \to 2$) and then from that partition to the global optimum ($2 \to 3$).}
	\label{fig:malaria_neo_1}
\end{figure*}

\begin{figure*}[ht]
	\includegraphics[width=\textwidth]{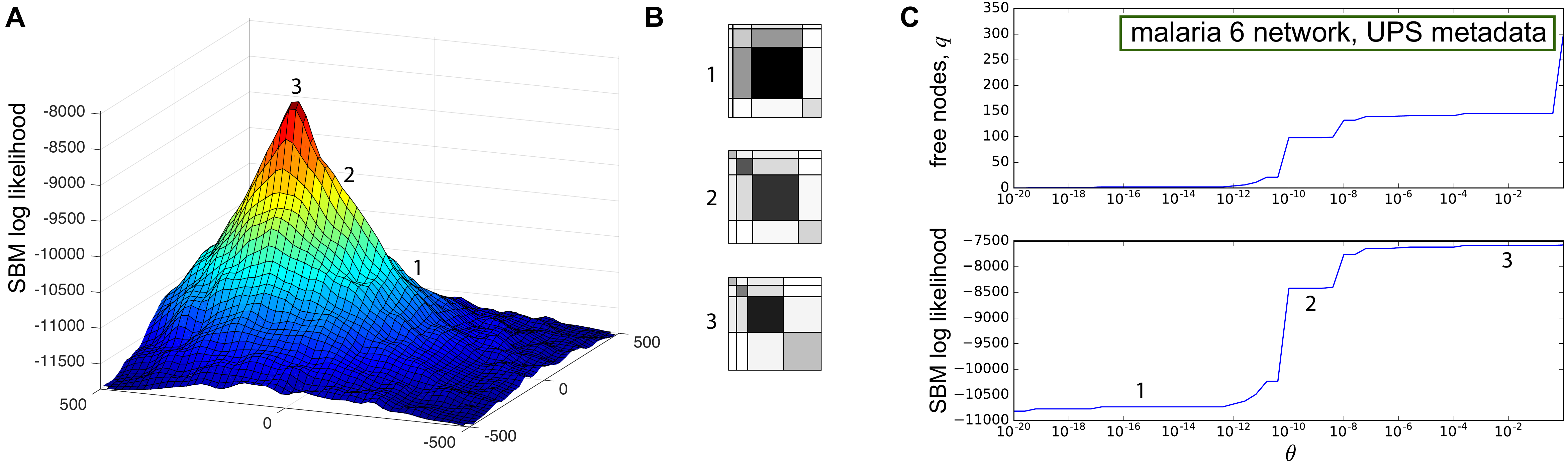}
	\caption{Results of the neoSBM on the malaria {\it var} gene network at locus six (``malaria 6'') using UPS metadata. (A) The SBM likelihood surface shows one prominent peak at the globally optimal partition. (B) Block density diagrams depict community structure for metadata and locally optimal partitions where darker color indicates higher probability of interaction. (C) The neoSBM, beginning from UPS metadata, interpolates between metadata 1 and the globally optimal SBM partition, traversing a local optimum during its path. The number of free nodes $q$ and SBM log likelihood as a function of $\theta$ shows two discontinuous jumps as the neoSBM traverses from the metadata to the locally optimal partition ($1 \to 2$), from that partition to another the global optimum ($2 \to 3$).}
	\label{fig:malaria_neo_6}
\end{figure*}

The metadata corresponding to upstream promoter sequence (UPS) are known to correlate with community structure in the malaria {\it var} gene networks, particularly at loci one and six \cite{larremore2013network,newman2016structure}. We provided the neoSBM with UPS metadata ($K=4$) and investigated the path of partitions between the metadata partition and the globally optimal partitions for each of the two networks.  Figures~\ref{fig:malaria_neo_1} (locus one) and \ref{fig:malaria_neo_6} (locus six) show likelihood surfaces, block density diagrams, and the neoSBM's outputs $q$ (free nodes) and SBM log likelihood.

Comparison of the neoSBM results for the same metadata on two different network layers reveals not only that the intermediate paths of locally optimal partitions differ but that the UPS metadata are more locally stable for the locus six network. This is indicated by the substantially larger value of $\theta$ at which the neoSBM switches from the metadata partition to the first intermediate local optimum. These transitions $1 \to 2$ involve different numbers of free nodes, however, indicating that the switch from optimum 1 to optimum 2 was accompanied by a much larger change in node mobility for the locus six network. Note that the neoSBM provides a more nuanced view of the relationship between UPS metadata and malaria layers one and six than the BESTest did, which found that UPS metadata were significantly correlated with the structures of both networks. 

\subsection{Synthetic network generation for the neoSBM}
\label{app_synth}
\begin{figure}[ht]
	\centering
	\includegraphics[width=1.0\columnwidth]{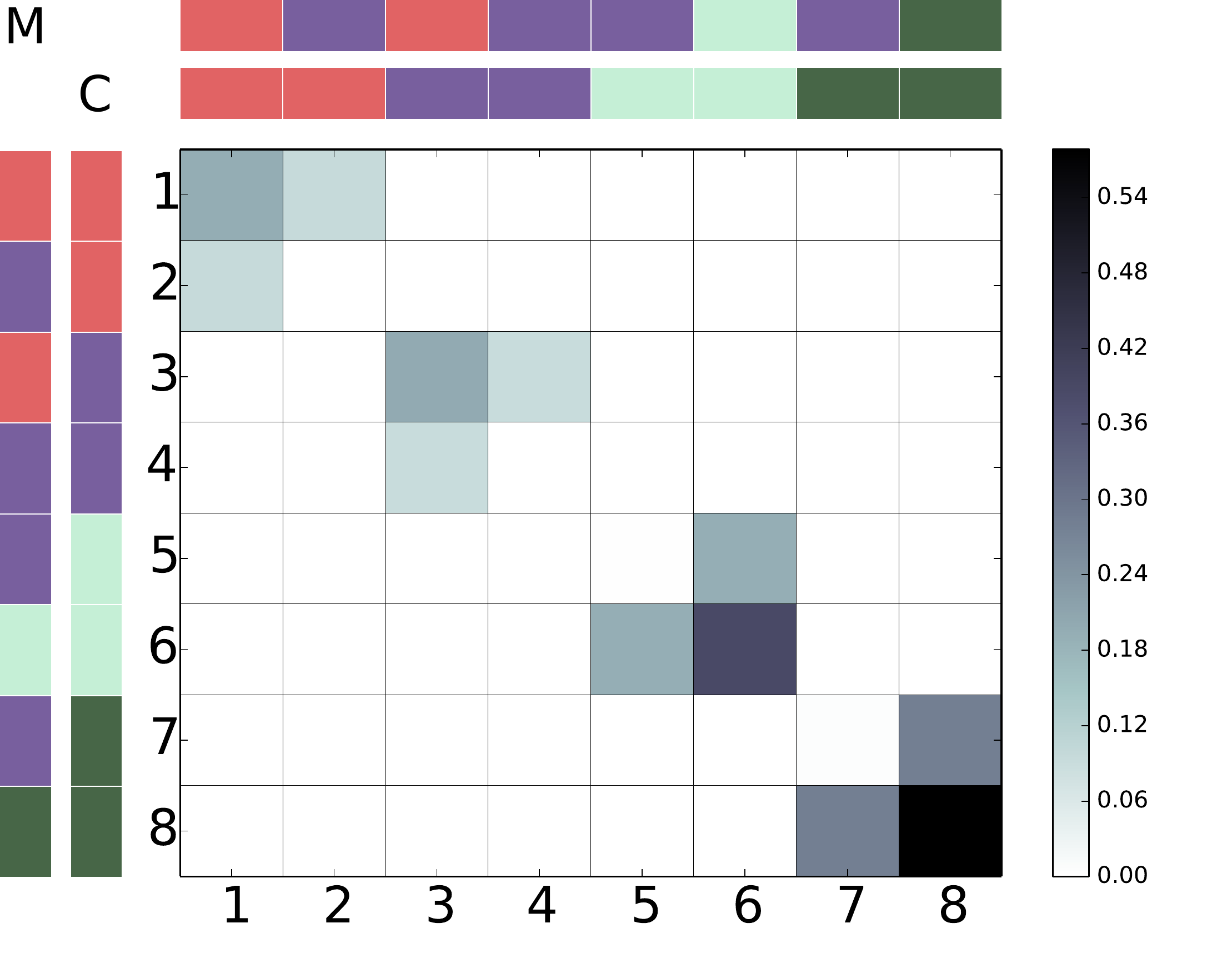}
	\caption{The block interaction matrix used to generate synthetic networks. The external colored rows and columns indicate the partition used as metadata (M) and the maximum likelihood partition under the SBM (C).}
	\label{fig_synth_model}
\end{figure}

The test that demonstrated the function of the neoSBM on synthetic data, depicted in Fig.~\ref{fig_synth_neo} of the main text, required networks with multiple local optima under the SBM: one corresponding to the inferred partition (global optimum) and at least one other to represent a relevant metadata partition. To create such a network, we divided vertices into $2K$ groups to create $K$ assortative communities, each of which was subdivided to contain a core and a periphery group. For $K=4$, Figure~\ref{fig_synth_model} shows the 8-block interaction matrix used to create the synthetic networks. By subsequently varying the mean degree within each block, we obtained two uncorrelated partitions when $K=4$, both of which are relevant to the network structure. Finally, we assigned as metadata the core-periphery structure containing one periphery group ($\{2,4,5,7\}$ in Fig.~\ref{fig_synth_model}) and three core groups (\{1,3\},\{6\},\{8\} in Fig.~\ref{fig_synth_model}). The partition inferred by the SBM in the absence of the neoSBM's likelihood penalty corresponds to the assortative group structure. 

\begin{minipage}{\linewidth}\noindent\matrixquote{\textbf{Morpheus}: \textit{Have you ever had a dream, Neo, that you seemed so sure it was real? But if were unable to wake up from that dream, how would you tell the difference between the dream world and the real world?}~\cite{matrix1999}}
\end{minipage}

\clearpage
\section{Blockmodel Entropy Significance Test}\label{supp:BESTest}

\noindent \matrixquote{{\bf Morpheus}: \textit{I'm trying to free your mind, Neo. But I can only show you the door. You're the one that has to walk through it.}~\cite{matrix1999}}

This Supplementary Text is divided into six subsections providing additional details on the blockmodel entropy significance test. 
\begin{itemize}
	\item Subsection \ref{Best} describes maximum likelihood parameter estimation for the SBM (I.a) and degree-corrected SBM (I.b). 
	\item Subsection \ref{Brapid} describes rapid computation of the entropy $H(\mathcal{G}; \mathcal{M})$ for the Bernoulli SBM and Multinomial degree-corrected SBM (DCSBM).
	\item Subsection \ref{Blike} demonstrates the mathematical link between our formulation of the SBM entropy and the SBM log likelihood which has been derived elsewhere \cite{karrer2011stochastic,peixoto2012entropy}. 
	\item Subsection \ref{Bmod} discusses the use of non-generative models like modularity.
	\item Subsection \ref{Bsynth} provides details on the generation of synthetic networks for the tests shown in Fig.~\ref{fig-BESTest_sensitivity}.
	\item Subsection \ref{Bapp} provides additional examples of results of the blockmodel entropy significance test using multiple different network data and metadata sets (see Supplementary Text \ref{supp:data}) as well as three additional generative network models beyond the SBM.
\end{itemize}

For convenience, we provide a reference table of notation used in derivations in this Supplementary Text.
\begin{table}[h!]
 \caption{Notation used in this Supplementary Text}
  \scriptsize
 \begin{center}
 \begin{tabular}{c|l} 
 \hline
Variable & Definition \\
 \hline
$\mathcal{G}$ & a network, $\mathcal{G}=(V,E)$  \\
$N$ & the number of nodes $\vert V \vert$\\
$\pi$ & a partition of nodes into groups \\
$K$ & the total number of groups \\
$\pi_i$ & the group assignment of node $i$ \\
$n_r$ & the number of nodes in group $r$ \\
$m_{rs}$ & the number of edges between groups $r$ and $s$ \\
$\kappa_{r}$ & the total degrees of group $r$, $\kappa_r = \sum_{s} m_{rs}$\\
$k_i$ & the degree of node $i$. \\
\hline
$H_{X}(\mathcal{G}\vert \mathcal{\pi})$ & entropy $H$ of model $X$ estimated for graph $\mathcal{G}$ using partition $\pi$\\
$\hat a$ & maximum likelihood estimate of model parameter $a$\\
$p_{ij}$ & the probability that an edge exists between nodes $i$ and $j$\\
 \hline
 \end{tabular}
 \end{center}
\end{table}

\subsection{Estimation of SBM parameters}\label{Best}

\subsubsection{Bernoulli SBM parameters}
Let the $N$ nodes of a network $\mathcal{G}$ be partitioned into $K$ groups, with the group assignment of node $i$ given by $\pi_i$. In the SBM, the probability of a link existing between any two nodes $i$ and $j$ depends only on the group assignments $\pi_i$ and $\pi_j$. This means that the entire model can be parameterized by a $K \times K$ matrix of block-to-block edge probabilities, $\omega$. Accordingly, let ${\bf \omega}$ be a matrix such that $p_{ij} = \omega_{\pi_i \pi_j}$ is the probability of a link existing between $i$ and $j$. Letting the number of nodes in group $r$ be $n_r$, then between two groups $r$ and $s$ there are $n_r n_s$ possible links, each of which has the same probability of existence, $\omega_{rs}$. This implies that the existence of the $n_r n_s$ edges between groups $r$ and $s$ will be determined by $n_r n_s$ independent Bernoulli trials, each with parameter $\omega_{rs}$. 

\begin{figure}[t]
	\centering
	\includegraphics[width=1.0\columnwidth]{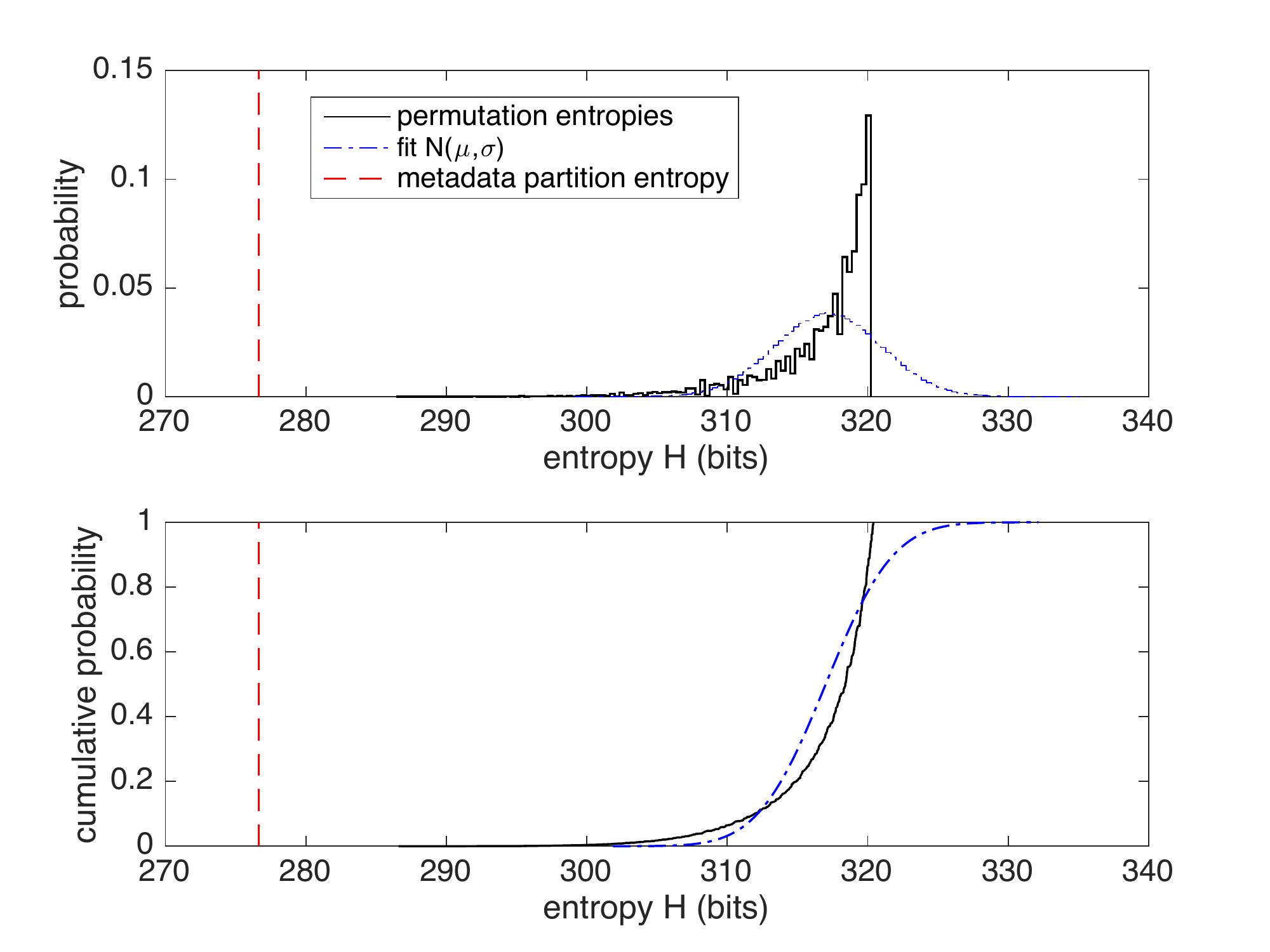}
	\caption{{\bf Distributions of permuted partition entropies are negatively skewed.} Probability density functions (top) and cumulative distribution functions (bottom) are shown for the entropies of partitions of the Karate Club network and its faction metadata. The red broken line indicates the point entropy of the metadata partition while the black solid line shows the distribution of entropies for $10^4$ independent permutations of the metadata partition. Note that these permutation entropies are far from normal; a normal distribution with equivalent mean $\mu$ and variance $\sigma^2$ is shown in blue for contrast.} 
	\label{fig-BESTest}
\end{figure}

We must now estimate the value of $\omega_{rs}$ for a network $\mathcal{G}$ whose nodes have been divided according to their assignments in partition $\pi$. Of course, any $\omega$ whose entries are positive will have some non-zero probability of having generated the observed links in $\mathcal{G}$. However, here we choose the values of $\omega$ to be those that maximize the likelihood of observing $\mathcal{G}$. Specifically, observe that of the $n_{r}n_{s}$ Bernoulli trials, there are $m_{rs}$ actual edges in the graph, i.e., $m_{rs}$ trial successes. Therefore, the maximum likelihood estimate of $\omega_{rs}$ is simply $\hat \omega_{rs} = m_{rs} / n_{r}n_{s}$. Thus, $\hat p_{ij} = \hat \omega_{\pi_i \pi_j}$. 

\subsubsection{Poisson degree-corrected SBM parameters}

In the degree-corrected Poisson SBM \cite{karrer2011stochastic}, it is still assumed that each link exists independently of the others, with some specified probability given by a block connectivity matrix $\omega$. However, this model differs in two key ways from the Bernoulli SBM. First, rather than each edge existing with probability $p_{ij}$, Poisson SBMs state that the {\it expected} number of edges between nodes $i$ and $j$ is given by a parameter $q_{ij}$, with the actual number of edges drawn from a Poisson distribution with identical mean. For very small values of $q$, the probability of an edge existing is approximately $q$, and thus if the graph is sufficiently sparse, Poisson SBMs behave similarly to Bernoulli SBMs, despite the fact that they could, in principle, generate multigraphs. 

The second way in which this degree-corrected Poisson SBM differs from the Bernoulli SBM is that the parameters $q_{ij}$ are no longer identical across the set of all $i$ in group $r$ and all $j$ in group $s$, as they are in the uncorrected SBM. Now, each node has a degree affinity $\theta_i$ so that $q_{ij} = \theta_{i} \theta_{j} e_{\pi_{i}\pi_{j}}$, where $e_{rs}$ is the $K \times K$ block structure matrix, controlling the numbers of links between groups, similar in principle to $\omega_{rs}$ above. The new parameters, $\theta_i$, properly chosen \cite{karrer2011stochastic}, can be used to specify the expected degree of each node.

As above, since we are given a network $\mathcal{G}$ and a fixed partition $\pi$, we must estimate the entries of $e$, as well as the values of $\theta$. The parameters can again be chosen to maximize the likelihood of observing $\mathcal{G}$, which are derived in \cite{karrer2011stochastic} but we do not derive here. First, $\hat e_{rs} = m_{rs}$, where $m_{rs} = \sum_{ij} A_{ij}\delta_{r,\pi_i}\delta_{s,\pi_j}$ is the number of links between groups $r$ and $s$ (or twice the number of links if $r=s$). Then, $\hat \theta_{i} = k_{i} / \kappa_{\pi_i}$, where $\kappa_r$ is the number of degrees connecting to group $r$, $\kappa_r = \sum_{s} m_{rs}$. Thus, $\hat q_{ij} = k_{i}k_{j}m_{\pi_{i}\pi_{j}}/\kappa_{\pi_i} \kappa_{\pi_j}$. We note that this maximum likelihood estimate is only valid in the regime that $k_{i}k_{j}m_{\pi_{i}\pi_{j}} \ll \kappa_{\pi_i} \kappa_{\pi_j}$.

\subsection{Rapidly computing entropy}\label{Brapid}

\subsubsection{Rapid Bernoulli SBM entropy}

Under either a Bernoulli-type SBM, a link exists between nodes $i$ and $j$ with probability $p_{ij}$, independently of all other links. This amounts to a Bernoulli trial or flip of a biased coin, and the entropy of this Bernoulli trial with parameter $p_{ij}$ is simply
\begin{equation}
	h(p_{ij}) \equiv - p_{ij} \log_2{p_{ij}} - (1-p_{ij}) \log_2{(1-p_{ij})}\,.
	\label{bernoulli}
\end{equation}
Hereafter, we will write simply $\log$ in place of $\log_2$. Because the Bernoulli trial on each link is conditionally independent of other links, the entropy of the network is the sum of all valid $h(p_{ij})$. For an undirected network this is
\begin{equation}
	H_{\text{SBM}}(\mathcal{G}) = \sum_{i \leq j} h(p_{ij}) = \frac{1}{2} \left [ \sum_{ij} h(p_{ij}) + \sum_{i} h(p_{ii}) \right ] \ .
	\label{teststatistic1}
\end{equation}

Under the SBM, the probabilities within each block are identical so we may group them and change to an index over groups, rewriting Eq.~\eqref{teststatistic1} as
\begin{equation}
	\displaystyle H_{\text{SBM}}(\mathcal{G}) = \frac{1}{2} \left [ \sum_{r s} n_{r}n_{s}h(\omega_{rs}) + \sum_{r} n_{r}h(\omega_{rr}) \right] \,.
	\label{teststatistic2}
\end{equation}
which may be simplified by plugging in the maximum likelihood estimate of $\hat \omega_{rs}$ and the definition of Bernoulli entropy $h$ Eq.~\eqref{bernoulli}, yielding
\begin{align}
	\displaystyle & H_{\text{SBM}}(\mathcal{G}) = \dots \nonumber \\
	 & -\frac{1}{2} \left [ \sum_{r s} m_{rs} \log \hat \omega_{rs}  + (n_r n_s - m_{rs}) \log (1 - \hat \omega_{rs} ) \right] + \mathcal{O}(n^{-1}) \,.
	\label{teststatistic3}
\end{align}
where we have noted that the diagonal terms are $\mathcal{O}(n^{-1})$ whenever $n_r = c n$ for some constant $c$.

Eq.~\eqref{teststatistic3} allows for a $\mathcal{O}(K^2)$ computation, rather than $\mathcal{O}(N^2)$ of Eq.~\eqref{teststatistic1}. For degree-corrected Bernoulli SBMs, entropies may be summed as in Eq.~\eqref{teststatistic1}, even though the rapid computation of Eq.~\eqref{teststatistic3} will not be valid. However, in what follows, we show the connection between model entropy $H$ and model log likelihood $\mathcal{L}$.

\subsubsection{Rapid Multinomial DCSBM entropy}
 
The degree-corrected SBM, introduced as a Poisson DCSBM by Karrer and Newman \cite{karrer2011stochastic}, can also be written in a ``Multinomial'' form in which each of the $m$ edges is placed sequentially, according to the multinomial probabilities $p_{ij}$ \cite{parkkinen2009block}. The values of $p_{ij}$ are defined as
\begin{equation}
	p_{ij} = \theta_{ir} \omega_{rs} \theta_{js} = \frac{k_i k_j e_{rs}}{2 m e_r e_s}
	\label{eq:multinomialp}
\end{equation}
where $\theta_{ir} = k_i / e_r$ if node $i$ is in group $r$, and 0 otherwise, and $\omega_{rs} = e_{rs}/2m$. Note that by definition, $\sum_{ij} p_{ij} = 1$. When constructing a network, $m$ edges are placed among the possible edge locations, with each one independently according to a categorical distribution with probabilities $p_{ij}$ \cite{parkkinen2009block}.

Since it is possible that multiple edges are formed between pairs of vertices, the entropy of this ensemble is not the entropy of $m$ categorical distributions with parameters $p$, but rather the entropy of the multinomial distribution with $m$ draws and $b$ ``bins'' with parameters $p$. Note that if there is a nonzero possibility of an edge between each pair of vertices, then $b=\binom{N}{2}$ [or $b=N^2$ in the directed case]. (There may be fewer than $\binom{N}{2}$ bins in the undirected case if some values of $e_{rs}$ are equal to 0, and similarly, there may be fewer than $N^2$ bins in the directed case if some values of $e{rs}$, $k_i^{\text{out}}$, or $k_i^{\text{in}}$ are equal to 0.) There is no closed-form expression for the entropy of a multinomial distribution but an accurate approximation has been derived in \cite{cichon2012bernoulli}, into which we substitute the parameters of the multinomial DSCBM, yielding
\begin{align}
	H = &\frac{1}{2} \log \left [ \left( 2 \pi m e \right)^{b-1} \prod_{ij; p_{ij} \neq 0} p_{ij} \right ] \dots \nonumber\\ 
	+&\frac{1}{12m}\left[ 3b - 2 - \sum_{ij; p_{ij} \neq 0} \frac{1}{p_{ij}} \right ] + \mathcal{O}\left( \frac{1}{m^2} \right )\ .
	\label{eq:multinomialentropy}
\end{align}
Thus, computing the entropy of this degree-corrected model \cite{parkkinen2009block} amounts to the rapid estimation of the parameters $p$ from Eq.~\eqref{eq:multinomialp} followed by computation of the entropy from Eq.~\eqref{eq:multinomialentropy}.

\subsection{Connecting entropy and log likelihood}\label{Blike}

The connection between model entropy $H$ and model log likelihood $\mathcal{L}$ enables the blockmodel entropy significance test to be expanded beyond the simple Bernoulli SBM to degree-corrected SBMs, Poisson SBMs, mixed-membership models, and other generative models with computable log likelihoods. 

We begin from Eq.~\eqref{teststatistic3} and use the Taylor series 
\begin{equation}
	(1-x) \ln(1-x) = -x+\sum_{\ell=2}^{\infty} \frac{x^{\ell}}{\ell(\ell-1)}, 
\end{equation}
in which we substitute $x = \hat \omega_{rs} = m_{rs} / n_r n_s$ to write Eq.~\eqref{teststatistic3} to leading order as
\begin{align}
	H_{\text{SBM}}(\mathcal{G}) \approx &-\frac{1}{2} \sum_{rs}  \bigg [m_{rs} \ln{ \left ( \frac{m_{rs}}{n_{r}n_{s}} \right )} - m_{rs} \dots \nonumber \\
	 &+ n_{r}n_{s}\sum_{\ell=2}^{\infty} \frac{1}{\ell(\ell-1)}\left ( \frac{m_{rs}}{n_{r}n_{s}} \right )^{\ell} \bigg ].
\end{align}

Finally, we note that $\frac{1}{2} \sum_{rs} m_{rs}$ is simply $|E|$, the total number of links in the network and therefore 
\begin{align}
	H_{\text{SBM}}(\mathcal{G}) \approx & |E|-\frac{1}{2} \sum_{rs} \bigg [m_{rs} \ln{ \left ( \frac{m_{rs}}{n_{r}n_{s}} \right )} \dots \nonumber \\
	& + n_{r}n_{s}\sum_{\ell=2}^{\infty} \frac{1}{\ell(\ell-1)}\left ( \frac{m_{rs}}{n_{r}n_{s}} \right )^{\ell} \bigg ].
	\label{teststatistic}
\end{align}
If all blocks of links are sparse, then $m_{rs} \ll n_{r}n_{s}$ and the terms in the infinite sum decay rapidly, leading to the first order approximation
\begin{equation}
	H_{\text{SBM}}(\mathcal{G}) \approx |E|-\frac{1}{2} \sum_{rs} m_{rs} \ln{ \left ( \frac{m_{rs}}{n_{r}n_{s}} \right )}.
	\label{teststatistic_sparse}
\end{equation}

Here we derived Eq.~\eqref{teststatistic} and Eq.~\eqref{teststatistic_sparse} by considering the conditionally independent entropies associated with every link of the network. However, the same equations can also be derived by calculating the size $\Omega$ of the ensemble of networks associated with the same SBM, and then taking a logarithm, $H = \log \Omega$. The log likelihood is the logarithm of the probability of observing an individual network realization from the ensemble, $\mathcal{L} = \log P$, and under the assumption that each graph in the ensemble occurs with the same probability, $P = 1/ \Omega$. Therefore, the entropy $H$ and the log likelihood $\mathcal{L}$ are related simply by $\mathcal{L} = -H$ \cite{peixoto2012entropy}. 

The relationship between the ``microcanonical'' entropy and log likelihood allows for the Blockmodel Entropy Significance test to be expanded easily to any generative model for networks for which a likelihood is easily computed,
\begin{equation}
	p\textrm{-value} = \text{Pr}\left [ \mathcal{L}(\mathcal{G};\tilde{\pi}) \geq \mathcal{L}(\mathcal{G};\mathcal{M}) \right ]\,.
	\label{BESTest-likelihood}
\end{equation} 

The Bernoulli SBM entropy Eq.~\eqref{teststatistic3} or its approximation for sparse networks Eq.~\eqref{teststatistic_sparse} are convenient because they are fast to compute---one need only to count links between groups, sizes of groups, and compute $\mathcal{O}(K^2)$ terms. By contrast, Eq.~\eqref{teststatistic1}, which is exact, requires $\mathcal{O}(N^2)$ computations. Depending on the assumptions involved, computing a log likelihood $\mathcal{L}$ may be more or less rapid, or more or less exact. 

In the additional tests in this Supplementary Text, we employ the Likelihood equations to apply the BESTest using the Poisson SBM and degree-corrected SBM, and use the rapid entropy equations for the Bernoulli SBM and Multinomial DCSBM.

Finally, we note that an alternative version of entropy that is not based on the blockmodel but instead by the size of the ensemble of networks with identical degree sequence and communities is discussed in Ref.~\cite{bianconi2009assessing}.

\subsection{Application of the significance test approach to non-generative models for community structure}\label{Bmod}

The blockmodel entropy significance test provides an estimate of how often a given partition provides a lower-entropy explanation of the data, as viewed through a particular model. While we have, so far, derived expressions for this test in terms of the entropy of a model Eq.~\eqref{eq-Q} or its likelihood Eq.~\eqref{BESTest-likelihood}, there exist many other approaches to community detection that are not generative, and therefore have neither a likelihood or an entropy. These models rely on a quality function or Hamiltonian which is optimized over partitions. Supposing that optimization of the Hamiltonian $\mathcal{Q}$ involves maximization, the test statistic is
\begin{equation}
	p\textrm{-value} = \text{Pr}\left [ \mathcal{Q}(\mathcal{G};\tilde{\pi}) \geq \mathcal{Q}(\mathcal{G};\mathcal{M}) \right ]\,.
	\label{BESTest-Q}
\end{equation} 
If optimization involves minimization of $\mathcal{Q}$, the direction of the inequality above should be reversed.

Modularity \cite{girvan2002community}, one of the most popular quality functions used for community detection, serves as an instructive example of the blockmodel entropy significance test in two ways. First, it is a measure of the strength of the assortment of links into communities, but has no generative model. Indeed, sampling from the space of networks with a particular modularity NP-hard \cite{brandes2006maximizing}. Second, the modularity score itself defines community structure narrowly as assortative, and therefore networks with disassortative structures, which have significant $p$-values using any SBM as the test model, are likely to be found to have non-significant $p$-values when Eq.~\eqref{BESTest-Q} is used. This emphasizes both the versatility of the test statistic, as well as the differences between definitions of community structure---spanning generative and non-generative models alike.

It is worth noting that the value of modularity is asymptotically zero whenever assortative communities are uncorrelated with the partition at which it is being evaluated---indeed, the premise of modularity maximization is to find communities whose internal edges defy expectation based on this uncorrelated null model. Thus, if metadata provide a partition of the network, and modularity is found to be exactly zero, then from the perspective of the particular type of assortative structure defined by modularity, there is not a significant relationship between metadata and community structure. On the other hand, simply finding that modularity under a particular metadata partition $\mathcal{Q}(\mathcal{G},\mathcal{M})$ is non-zero need not imply that the relationship is or is not statistically significant in the sense of Eq.~\eqref{BESTest-Q}; the test must be performed.
 
\subsection{Generation of synthetic networks for blockmodel entropy significance test}\label{Bsynth}

The tests described in the main text, and detailed in this Supplementary Text, will yield a $p$-value which indicates the extent to which a set of metadata (and a generative model) describes a network better than a random partition. In order to understand the sensitivity of the BESTest, we generated sets of synthetic networks and synthetic metadata, applied the BESTest to them, and produced Fig. \ref{fig-BESTest_sensitivity}. Here we describe the process used to generated those synthetic networks. 

We generated networks of $N=1000$ nodes and two planted communities $r$ and $s$ using the (Bernoulli) SBM. Each node was assigned to one of the communities ($\mathcal{T}_i=r$ or $\mathcal{T}_i=s$) with equal probability. We then generated a network with a given community strength $\epsilon=\omega_{rs} / \omega_{rr}$ such that low values of $\epsilon$ generate strongly assortative communities with few connecting edges between them and as $\epsilon$ grows, the generated communities become weaker, producing a random graph with no communities when $\epsilon = 1$. For each node $i$, with probability $\ell$ we assigned its metadata label to be its community label ($\mathcal{M}_i=\mathcal{T}_i$), otherwise we assigned it a uniformly random label. Thus, as $\ell$ increases from $0$ to $1$ the metadata labels correlate more with the planted communities, and the probability that any individual node's metadata label matches its community label is $\ell(1) + (1-\ell)(1/2) = (1+\ell)/2$. 

\subsection{Additional applications of the BESTest to real data}\label{Bapp}

We now present and discuss the results of applying the BESTest to the Lazega Lawyers and Malaria data sets (see Supplementary Text \ref{supp:data}).

\subsubsection{Lazega Lawyers}
\begin{table}[t]
 \caption{Lazega Lawyers: BESTest p-values}
 \label{tab:pvals}
 \begin{center}
 \begin{tabular}{l|ccccccccc} 
 \hline
 & \multicolumn{5}{c}{Attribute} \\ 
 Network & Status & Gender & Office & Practice & Law School \\
 \hline
& \multicolumn{5}{c}{SBM}\\
Friendship	&$<10^{-6}$	&$0.034$	&$<10^{-6}$	&$0.033$	&$0.134$	\\
Cowork	&$<10^{-3}$	&$0.094$	&$<10^{-6}$	&$<10^{-6}$	&$0.922$	\\
Advice	&$<10^{-6}$	&$0.010$	&$<10^{-6}$	&$<10^{-6}$	&$0.205$	\\
 \hline
& \multicolumn{5}{c}{DCSBM}\\
Friendship	&$<10^{-6}$	&$0.001$	&$<10^{-6}$	&$0.002$	&$0.094$	\\
Cowork	&$<10^{-6}$	&$0.842$	&$<10^{-6}$	&$<10^{-6}$	&$0.938$	\\
Advice	&$<10^{-6}$	&$0.205$	&$<10^{-6}$	&$<10^{-6}$	&$0.328$	\\
\hline
& \multicolumn{5}{c}{Poisson SBM}\\
Friendship	&$<10^{-6}$	&$0.046$	&$<10^{-6}$	&$0.044$	&$0.167$	\\
Cowork	&$<10^{-3}$	&$0.099$	&$<10^{-6}$	&$<10^{-6}$	&$0.977$	\\
Advice	&$<10^{-6}$	&$0.013$	&$<10^{-6}$	&$<10^{-6}$	&$0.316$	\\
 \hline
& \multicolumn{5}{c}{Poisson DCSBM}\\
Friendship	&$<10^{-6}$	&$<10^{-3}$	&$<10^{-6}$	&$<10^{-3}$	&$0.014$	\\
Cowork	&$<10^{-4}$	&$0.969$	&$<10^{-6}$	&$<10^{-6}$	&$0.781$	\\
Advice	&$<10^{-5}$	&$0.018$	&$<10^{-6}$	&$<10^{-6}$	&$0.046$	\\
 \hline
 \end{tabular}
 \end{center}
\end{table}

We applied the BESTest to all three Lazega Lawyers networks (Friendship, Cowork, Advice) which share the same set of nodes but have different sets of edges, representing different relationships between individuals.  There were five sets of node metadata (Status, Gender, Office, Practice, and Law School). We applied the BESTest to each combination of network and metadata, using four generative models (SBM, degree-corrected SBM, Poisson SBM, and Poisson degree-corrected SBM). These results are shown in Table \ref{tab:pvals}.

First, note that values between Bernoulli and Poisson models are not identical, though they are similar, implying that the models are not entirely interchangeable. More importantly, however, the results for degree-corrected and degree-uncorrected models are substantially more different, with relationships varying from significant under one model to insignificant under another. This highlights the fact that metadata can explain patterns of group structure in a network only through the lens of a particular network generative model; a change in the model may impact the metadata's ability to explain patterns in network community structure. 

Second, note that under all models, for each network there exist multiple sets of metadata that are significant. Similarly, there exist multiple networks for which any individual set of metadata is significant. This fundamentally undermines the notion that one should expect a single set of metadata to function as ground truth, given that multiple sets of metadata explain multiple networks. 

\subsubsection{Malaria}

We applied the BESTest to nine layers of a network of malaria parasite genes (Malaria 1-9) using four generative models (SBM, degree-corrected SBM, Poisson SBM, and Poisson degree-corrected SBM). Three sets of metadata exist for these networks, (parasite origin, CP group, and UPS), described in detail in Supplementary Text \ref{supp:data}. 

The {\it parasite origin} results are shown in Table \ref{tab:mal_pvals}, and none of the $p$-values listed is significant. This result indicates that when the nodes of each layer are divided into groups based on parasite origin, the entropy of the resulting model is no better than assigning the nodes to groups at random. This implies, in turn, that the malaria parasite antigen genes do not group by the parasite from which they came, confirming previous observations \cite{larremore2013network}. However, as shown in Fig.~\ref{fig-BESTest_sensitivity} the BESTest is sensitive to even small levels of explanatory power provided by metadata, indicating that parasite origin has truly no bearing on the community structure of malaria parasite antigen genes, for all four generative models tested. 

\begin{table}[t]
 \caption{Malaria: BESTest p-values for parasite origin metadata}
 \label{tab:mal_pvals}
 \begin{center}
 \begin{tabular}{l|ccccccccc} 
 \hline
 & \multicolumn{4}{c}{Model} \\ 
 Network & SBM & DCSBM & Poi. SBM & Poi. DCSBM \\
 \hline
Malaria 1	&$0.566$	&$0.096$	&$0.606$	&$0.086$	\\
Malaria 2	&$0.064$	&$0.148$	&$0.066$	&$0.143$	\\
Malaria 3	&$0.536$	&$0.389$	&$0.532$	&$0.501$	\\
Malaria 4	&$0.588$	&$0.617$	&$0.604$	&$0.644$	\\
Malaria 5	&$0.382$	&$0.077$	&$0.369$	&$0.087$	\\
Malaria 6	&$0.275$	&$0.923$	&$0.293$	&$0.751$	\\
Malaria 7	&$0.020$	&$0.388$	&$0.019$	&$0.501$	\\
Malaria 8	&$0.464$	&$0.176$	&$0.468$	&$0.172$	\\
Malaria 9	&$0.115$	&$0.067$	&$0.108$	&$0.200$	\\
 \hline
 \end{tabular}
 \end{center}
\end{table}

On the other hand, it is known that the genes represented by the nodes of the malaria parasite networks are correlated with CP group and UPS metadata. As shown in Tables \ref{tab:mal_CP} and \ref{tab:mal_UPS} the BESTest indeed finds that this is the case, with a handful of exceptions, again confirming previous results that used less sophisticated techniques \cite{larremore2013network}.

\begin{table}[h]
 \caption{Malaria: BESTest p-values for CP group metadata}
 \label{tab:mal_CP}
 \begin{center}
 \begin{tabular}{l|ccccccccc} 
 \hline
 & \multicolumn{4}{c}{Model} \\ 
 Network & SBM & DCSBM & Poi. SBM & Poi. DCSBM \\
 \hline
Malaria 1	&$<10^{-5}$	&$<10^{-5}$	&$<10^{-5}$	&$<10^{-5}$	\\
Malaria 2	&$<10^{-5}$	&$<10^{-5}$	&$<10^{-5}$	&$<10^{-5}$	\\
Malaria 3	&$<10^{-5}$	&$<10^{-5}$	&$<10^{-5}$	&$<10^{-5}$	\\
Malaria 4	&$<10^{-5}$	&$<10^{-5}$	&$<10^{-5}$	&$<10^{-5}$	\\
Malaria 5	&$<10^{-5}$	&$<10^{-5}$	&$<10^{-5}$	&$<10^{-5}$	\\
Malaria 6	&$<10^{-5}$	&$<10^{-5}$	&$<10^{-5}$	&$<10^{-5}$	\\
Malaria 7	&$<10^{-5}$	&$<10^{-5}$	&$<10^{-5}$	&$<10^{-5}$	\\
Malaria 8	&$<10^{-5}$	&$<10^{-5}$	&$<10^{-5}$	&$<10^{-5}$	\\
Malaria 9	&$<10^{-5}$	&$<10^{-5}$	&$<10^{-5}$	&$<10^{-5}$	\\
 \hline
 \end{tabular}
 \end{center}
\end{table}
\begin{table}[h]
 \caption{Malaria: BESTest p-values for UPS metadata}
 \label{tab:mal_UPS}
 \begin{center}
 \begin{tabular}{l|ccccccccc} 
 \hline
 & \multicolumn{4}{c}{Model} \\ 
 Network & SBM & DCSBM & Poi. SBM & Poi. DCSBM \\
 \hline
Malaria 1	&$<10^{-5}$	&$<10^{-5}$	&$<10^{-5}$	&$<10^{-5}$	\\
Malaria 2	&$<10^{-5}$	&$<10^{-5}$	&$<10^{-5}$	&$<10^{-5}$	\\
Malaria 3	&$<10^{-5}$	&$<10^{-5}$	&$<10^{-5}$	&$<10^{-5}$	\\
Malaria 4	&$<10^{-5}$	&$<10^{-5}$	&$<10^{-5}$	&$<10^{-5}$	\\
Malaria 5	&$<10^{-5}$	&$<10^{-5}$	&$<10^{-5}$	&$<10^{-5}$	\\
Malaria 6	&$<10^{-5}$	&$<10^{-5}$	&$<10^{-5}$	&$<10^{-5}$	\\
Malaria 7	&$<10^{-5}$	&$<10^{-5}$	&$<10^{-5}$	&$<10^{-5}$	\\
Malaria 8	&$<10^{-5}$	&$<10^{-5}$	&$<10^{-5}$	&$<10^{-5}$	\\
Malaria 9	&$<10^{-5}$	&$<10^{-5}$	&$<10^{-5}$	&$<10^{-5}$	\\
 \hline
 \end{tabular}
 \end{center}
\end{table}

\noindent \matrixquote{{\bf Morpheus}: \textit{The Matrix is a system, Neo. That system is our enemy. But when you're inside, you look around, what do you see? $\dots$ The very minds of the people we are trying to save. But until we do, these people are still a part of that system and that makes them our enemy. You have to understand, most of these people are not ready to be unplugged. And many of them are so inured, so hopelessly dependent on the system, that they will fight to protect it.}~\cite{matrix1999}}

\clearpage
\section{No optimal community detection algorithm}\label{supp:bijection}

\noindent\matrixquote{{\bf Spoon boy}: \textit{Do not try and bend the spoon---that's impossible. Instead, only try to realize the truth.}\\
{\bf Neo}: \textit{What truth?}\\
{\bf Spoon boy}: \textit{There is no spoon.}\\
{\bf Neo}: \textit{There is no spoon?}\\
{\bf Spoon boy}: \textit{Then you will see that it is not the spoon that bends, it is only yourself.}~\cite{matrix1999}
}

In the main text we argue that that the goal of recovering ground truth communities is ill posed because it lacks a unique solution and we also claim a ``No Free Lunch'' theorem for community detection. In this Supplementary Text, we describe and expound those claims using technical arguments. 

For convenience, we provide a reference table of notation used in derivations in this Supplementary Text.

\begin{table}[h!]
	\caption{Notation used in this Supplementary Text}
	 \scriptsize
	\begin{center}
	\begin{tabular}{c|l} 
	\hline
	Variable & Definition \\
	\hline
	$\mathcal{G}$ & a network, $\mathcal{G}=(V,E)$  \\
	$N$ & the number of nodes $|V|$ \\
	$\mathcal{T}$ & ground truth (planted) partition\\
	$\mathcal{C}$ & detected communities partition\\
	$g$ & generative model, maps a partition to a network. $\mathcal{G}=g(\mathcal{T})$ \\
	$f$ & comm. detection method, maps $\mathcal{G}$ to a partition $\mathcal{C} = f(\mathcal{G})$\\
	$\ell(\cdot, \cdot)$ & an error or loss function, returns a scalar\\
	$X$ & the space of possible inputs, i.e., all possible graphs $\mathcal{G}$ \\ 
	$Y$ & the space of possible outputs, i.e., all possible partitions \\
	$\gamma$ & the true relationship between $X$ and $Y$ \\
	$h$ & the hypothesis about the relationship between $X$ and $Y$ \\
	$\sigma_X$ & probability density over $X$ \\
	$\Lambda(\ell)$ & total loss across all possible inputs for loss function $\ell$ \\ 
	$u, v$ & two partitions of $N$ objects\\ 
	$\Omega$ & the set of all possible partitions of $N$ objects. \\ 
	$\mathcal{B}_N$ & the $N$th Bell number, the number of possible ways\\ 
	& $\quad$ that $N$ objects can be partitioned. $\mathcal{B} = \lvert \Omega \rvert$\\ 
	\hline
	\end{tabular}
	\end{center}
\end{table}
\vspace{-0.1in}

\subsection{Ground-truth community detection is an ill-posed inverse problem}

A problem that is well posed satisfies three properties: (i) a solution exists, (ii) the solution is unique, and (iii) the solution's behavior changes continuously with the problem's initial conditions. The classic example of an ill-posed problem is the inverse heat equation, which violates condition (iii) because its solution (the distribution of temperature in the past) is highly sensitive to changes in the distribution of temperatures at the present. The problem of reproducing ground truth communities $\mathcal{T}$ from a network $\mathcal{G}$ by formulating the correct community detection algorithm $f^*$ is ill posed because it fails condition (ii), i.e., community detection has no unique solution. 

\vspace{0.1in}
\begin{addmargin}[3em]{3em}
\noindent{\bf Definition}: The {\it ground truth community detection problem}: given a fixed network $\mathcal{G}$ in which there has been hidden some ground truth partition $\mathcal{T}$, find the true communities that were planted in, embedded in, or used to generate the network. In other words, given $\mathcal{G}$, find the $\mathcal{T}$ such that $\mathcal{G} = g(\mathcal{T})$.
\end{addmargin}
\vspace{0.05in}

We now argue that the ground truth community detection problem is ill posed because its solution is not unique. The intuition behind this argument is that any network $\mathcal{G}$ could have been created using many different planted partitions via different generative processes. Therefore, searching for the ground truth partition without knowing the exact generative mechanism is an impossible task; there is no ground truth.

\vspace{0.05in}
\begin{addmargin}[3em]{3em}
\noindent{\bf Theorem 1}: For a fixed network $\mathcal{G}$, the solution to the ground truth community detection problem is not unique.
\end{addmargin}

\vspace{0.05in}
\noindent{\it Proof:} We first show that the graph $\mathcal{G}$ can be produced by using two different planted partitions, $\mathcal{T}_1$ and $\mathcal{T}_2$ with $\mathcal{T}_1 \neq \mathcal{T}_2$. Let $\mathcal{T}_1$ be the trivial 1-partition in which all vertices are in the same group, and let $g_1$ be the generative model of Erd\H os-R\'enyi random graphs with probability $p \in (0,1)$. Then the model and partition $g_1(\mathcal{T}_1)$ create $\mathcal{G}$ with non-zero probability. Let $\mathcal{T}_2$ be the trivial $N$-partition in which each vertex is in its own group, and let $g_2$ be a generative model that specifies the exact number of edges between all groups, such that $g_2(\mathcal{T}_2)$ produces $\mathcal{G}$ with probability one. We therefore have two partitions $\mathcal{T}_1 \neq \mathcal{T}_2$ and both $g_1(\mathcal{T}_1)$ and $g_2(\mathcal{T}_2)$ can create $\mathcal{G}$. Since two different planted partitions may be responsible for $\mathcal{G}$, both are potential solutions of the community detection problem. Therefore, the solution to the community detection problem is not unique for the network $\mathcal{G}$. To complete the proof, note that the $1$-partition and $N$-partition argument above applies equally well to any network $\mathcal{G}$. \qed
\vspace{0.05in}

The theorem above relies on two trivial partitions, the 1-partition and the $N$-partition in its proof, but other examples exist as well: consider the generative model $g_{\mathcal{G}^*}$ which maps any partition that it is given to some fixed network $ \mathcal{G}^*$, essentially ignoring the information provided by the partition [similar to case (i) in the main text]. These models, while sufficient for the proof, are not particularly interesting for practictioners, yet non-trivial models and partitions also exist for any network. For instance, the Karate Club network may have plausibly been produced by the SBM with a core-periphery partition or by the degree-corrected SBM with a social faction partition \cite{karrer2011stochastic}.
 
Belief in ground truth $\mathcal{T}$ necessitates a belief in a specific generative mechanism $g$ which together produced the network $\mathcal{G}$. For real-world networks, which may arise through more complex processes than those described here, we do not know the generative mechanism.   Theorem 1 means that, in these cases, it is impossible to recover the {\it true} partition because {\it any} partition may plausibly have been used to generate the network. Therefore the ground truth community detection problem is ill-posed for any network for which the generative process is unknown because there is no unique solution. Put differently, it is impossible to uniquely solve an inverse problem when the function to be inverted is not a bijection.

\subsection{No Free Lunch for community detection}
The ``no free lunch'' (NFL) theorem~\cite{wolpert1996lack} for machine learning states that for supervised learning problems, the expected misclassification rate, summed over all possible data sets, is independent of the algorithm used. In other words, averaged over all problems, every algorithm has the same performance. Therefore, if algorithm $f_1$ outperforms algorithm $f_2$ for one set of problems, then there exists some other set of problems for which algorithm $f_2$ outperforms algorithm $f_1$. In other words, it is impossible to get overall better performance without some cost; there is no free lunch.

The NFL theorem holds for community detection, and clustering problems in general. Demonstrating this requires that we first translate the community detection problem into the language and notation of the Extended Bayesian Framework (EBF) used in the NFL theorems for supervised learning. Then, under an appropriate choice of error (or ``loss'') function $\ell$, the performance of any community detection method $f$, summed over all problems $\{ g, \mathcal{T} \}$, is identical
\begin{equation}
	\sum_{g,\mathcal{T}} \ell\big(\mathcal{T},f\left(g(\mathcal{T})\right)\big) = \Lambda(\ell) \quad \forall f \enspace ,
	\label{eq_NFL}
\end{equation}
where $\Lambda(\ell)$ depends on the particular error function $\ell$ but is otherwise a constant, representing the total error.

In the following, we map community detection notation to EBF notation, provide a guiding example, and then resolve a subtle issue related to the loss function $\ell$. We then discuss the implications of this result for future studies of community detection. The proofs of the NFL theorems are not recapitulated here, but are fully detailed in Ref.~\cite{wolpert1996lack} and discussed extensively elsewhere.

\subsubsection{Community detection in the Extended Bayesian Framework}

The Extended Bayesian Framework (EBF) is a framework---a set of variables, definitions, and assumptions---for supervised learning that provides a clear and precise description of the problem. It is important in both the proof and implications of the NFL theorem, and was formalized at length in Ref.~\cite{wolpert1996lack}. In what follows, random variables will be denoted by capital letters, e.g. $X$, while instances of random variables will be denoted by the corresponding lowercase letters, e.g. $x$. In the EBF, we suppose that there exists an input space $X$, an output space $Y$, and that each of these has a countable (but possibly infinite) number of elements, $|X| = n$ and $|Y| = r$. The fundamental relationship to be learned is how $X$ and $Y$ are related, and to that end, let $\gamma$ be the true or target relationship between $X$ and $Y$, i.e., $\gamma$ is the conditional distribution of $Y$, given $X$. The points in the space $X$ need not be distributed uniformly either, so we also specify $\sigma$, the probability density function of points $x$ in the input space $X$, i.e., $P(x|\sigma) = \sigma_{X}$. In the nomenclature of community detection, the input $x \in X$ is simply the observed graph $\mathcal{G}$, and the output $y \in Y$ is the true partition into communities $\mathcal{T}$ for the nodes described by $x$. To solve a community detection problem, we hope to predict the true communities $y$ from the input graph $x$; a community detection method will be successful when its hypothesized relationship $h$ is an accurate representation of the true relationship $\gamma$ between $X$ and $Y$.

In supervised learning, for which the NFL theorems were originally proved, we aim to learn the relationship between $X$ and $Y$ from a training set $d$ which consists of $m$ ordered pairs of samples from $X$ and $Y$, $\{d_{X}(i),d_{Y}(i)\}_{i=1}^{m}$. In response to the training data, the learning algorithm produces a hypothesis $h$ in the form of an $x$-conditioned probability distribution over values $y$. The way in which the learning algorithm produces a hypothesis from training sets is described by $P(h|d)$, the distribution over hypotheses conditioned on the observed data. Note that the algorithm learns from the data alone and is independent of $\gamma$, i.e., $P(h|d,f) = P(h|d)$. If the algorithm performs well the hypothesis $h$ will have high correspondence with the true relationship $\gamma$. Therefore, in supervised learning, algorithms are evaluated by their ability to make sufficient use of a limited training set to provide good predictions of $y$ given $x$ {\it not} in the training set. On the other hand, in unsupervised learning---a category which includes clustering and community detection---the training set $d$ is empty ($m=0$), so the prediction $h$ is based solely on the prior beliefs encoded in the model $P(h)$. We note that in the NFL theorems for supervised learning, the independence of training data $d$ from $\gamma$ and $\sigma$ is important to establish, but for unsupervised tasks, the set $d$ is empty so it is trivially independent of $\gamma$ and $\sigma$. 

To better understand the EBF for community detection, an example is helpful. Consider the problem of finding two planted communities in a network $\mathcal{G}$. The true relationship $\gamma$ between the network and its partition is hidden. Given only $\mathcal{G}$---which is a point in the space of graphs $X$---fitting the parameters of an SBM, maximizing modularity, or using another method of our choice, produces a hypothesis $h$, which is a prediction about which nodes belong to which groups. If these communities are found correctly by the algorithm, then $h$ will be highly correlated with the true communities mapped by $\gamma$. (This is equally true for both hard partitions, where each node belongs to only one group, and soft partitions, where each node may be distributed over multiple groups.) In other words, $h$ estimates $\gamma$ based on a point in $X$ called $\mathcal{G}$. Because the estimate $h$ is based {\it only} on $\mathcal{G}$ and the assumptions of the algorithm $P(h)$, it reproduces $\gamma$ with possibly limited accuracy, and therefore its community assignments may or may not be highly correlated with the true assignments $\mathcal{T} \in Y$. Increasing the size of the input data set may help with accuracy as well: by generating a larger graph using the same generative model, $\mathcal{G}$ supplies a different point in $X$ providing more information to the community detection method. This may allow the estimate $h$ to produce better predictions of $\gamma$, thereby producing a more accurate partitioning of nodes into their true communities, but only if the model $P(h)$ is sufficiently aligned to reality $P(\gamma)$.
 
All learning algorithms make some prior assumptions, in the form of $P(h)$, about the possible relationships between inputs and outputs.  For unsupervised methods such as community detection, there is a much greater importance associated with these assumptions because they do not have access to training data. For instance, a supervised algorithm could supposedly start from a uniformly ignorant prior $P(h)$ and rely on having a sufficiently large training set that $P(h|d)$ is informative.  When there is no training data it is necessary that $P(h)$ is informative of the possible input-output relationship. Thus, community detection algorithms encode beliefs or definitions of community structure, and these beliefs constitute a prior over the kinds of problems that we expect to see. Some methods, for example, search only for assortative~\cite{girvan2002community,ball2011efficient} or disassortative~\cite{larremore2014efficiently} community structures, while other are more flexible and can find mixtures of assortative, disassortative, and core-periphery structures~\cite{holland1983stochastic, nowicki2001estimation, karrer2011stochastic, peixoto2014hierarchical} and allow for nodes to belong to multiple communities~\cite{ball2011efficient, airoldi2009mixed}. 

\subsubsection{Loss functions and \textit{a priori} superiority}

So far, we have discussed the phrasing of community detection in the language of EBF but have not described the way in which error (also called loss or cost) is measured. The error function quantifies the accuracy of predictions, and the EBF introduces a random variable $C$ which represents the error associated with a particular $\gamma$ and $h$, i.e., the error associated with using a particular algorithm for a particular problem. Conceptually, this is what the community detection literature attempts to estimate when algorithms are compared based on their ability to recover planted communities in synthetic data. More formally, $C$ is measured by the distribution $P(c|h,\gamma,d)$, which incorporates the relationships between the test set and the generating process, as well as the way in which the hypothesis is related to the training data. Therefore, the quantity of interest to those developing algorithms is the expected error, $E(C | h, \gamma, d)$. For example, in the context of supervised learning, choosing the loss function $\ell$ to be the average misclassification rate is common. For the purposes of community detection, misclassification rate is not of interest for a pedantic but important reason: for community detection and other related unsupervised tasks such as clustering, permutations of the group labels are inconsequential because the partition is the desired outcome; labeling two groups $a$ and $b$ is equivalent to labeling them $b$ and $a$. As a result, many of the loss functions typically used to compare partitions have a ``geometric'' structure that implies an {\it a priori} superiority of some algorithms, which would appear to contradict the NFL theorem. We now discuss one such loss function frequently used to evaluate community detection algorithms, the normalized mutual information, and the structure that it imposes on the space of partitions. 

\begin{figure}[t]
	\centering
	\includegraphics[width=0.5\textwidth]{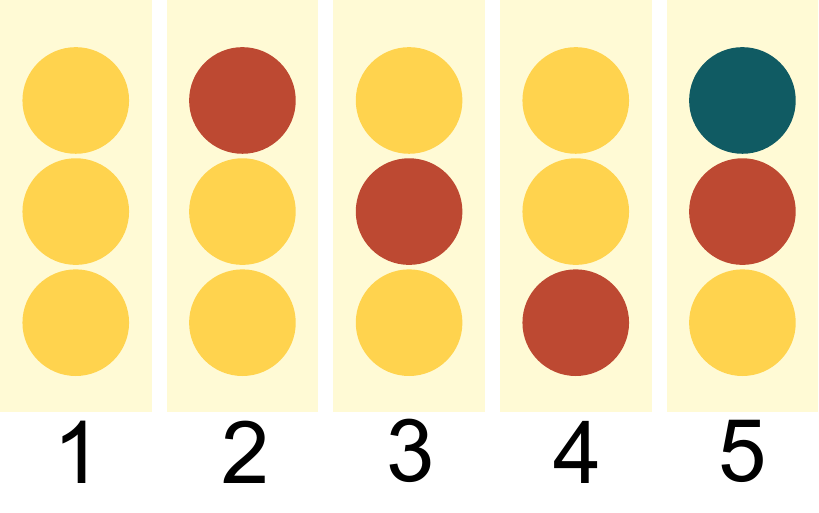}
	\caption{The five distinct ways to partition three nodes. Normalized mutual information and adjusted mutual information between each pair of partitions are presented in Tables~\ref{tab_nmi} and \ref{tab_ami}, respectively.}
	\label{fig_partitions}
\end{figure}

Normalized mutual information is an information-theoretic measurement of similarity between two partitions that treats both partitions as statistical objects. For a partition $u$ of $N$ objects into $K_u$ groups, the probability that an object chosen uniformly at random falls into group $u_i$ is $p_i = \lvert u_i \rvert / N$, $i = 1 \dots K_u$. The entropy associate with a partition $u$ is then the entropy of its corresponding distribution $p$, 
$$H(u) = - \sum_{i=1}^{K_u} p_i \log \left ( p_i \right )\ .$$
When comparing two partitions $u$ and $v$ of the same set of objects, each object belongs to some group $u_i$ in the first partitions and some other group $v_j$, $j = 1 \dots K_v$ in the second partition, with the corresponding probability $p_{ij}$. The mutual information between the two partitions is therefore
$$I(u,v) = \sum_{i=1}^{K_u} \sum_{j=1}^{K_v} p_{ij} \log \left (\frac{p_{ij}}{p_{i} p_{j}} \right )\ ,$$
which can be normalized to define normalized mutual information as
\begin{equation}
	\text{NMI}(u,v) = \frac{I(u,v)}{\sqrt{H(u) H(v)}}\ .
\end{equation}
Other normalizing factors in the denominator are possible, including $\frac{1}{2}[H(u)+H(v)]$ and $\text{max}\{H(u),H(v)\}$; see \cite{vinh2009information}. NMI maps partitions to the unit interval, with $0$ indicating that two partitions are uncorrelated and $1$ indicating that they are identical (even if the groups labels differ).  

To understand how an error function imposes a geometric structure, consider a simple problem (unrelated to community detection) of predicting, based on some inputs $X$, a point in the unit circle in $Y = \left \{ y\enspace \big | \enspace   \| y \| \leq 1,\enspace y \in \mathcal{R}^2 \right \}$. If all points in $Y$ are equally likely, then an algorithm that guesses the center of the circle $h=0$ will outperform an algorithm that guesses a point on the boundary $h \in \partial Y$, simply due to the fact that the center of the circle is, on average, closer to the other points of the circle than any boundary point. Normalized mutual information imposes a geometric structure on the space of partitions in a similar way.

Consider a loss function based on normalized mutual information (NMI) and imagine a community detection algorithm that entirely ignores the network and simply returns a fixed partition of the vertices. As in the example above, NMI provides a geometrical structure on the space of partitions, an algorithm that always returns a partition toward the middle of the space of partitions will outperform an algorithm that always returns a partition on the boundary of that space. To demonstrate this point, Fig.~\ref{fig_partitions} shows all five possible partitions of three vertices, and Table~\ref{tab_nmi} shows their NMI for all pairwise comparisons. Averaged over all possible correct answers, an algorithm that consistently predicts partition $5$ will outperform all others, and an algorithm that consistently predicts partition $1$ will underperform all others. However, this structure is a known issue of NMI, and so other error functions and corrections have been proposed such as the adjusted mutual information (AMI), which accounts for the geometry of the space~\cite{vinh2009information}. Table~\ref{tab_ami} shows the AMI for the same set of partitions, and the expected AMI is zero except for the partition that contains only a single group and the partition of each node into separate groups. In the case of these partitions, the $1$-partition and the $N$-partition, the expected AMI is the reciprocal of the Bell number $\mathcal{B}_N$---the Bell number is the total number of distinct ways that $N$ objects can be partitioned, and it grows superexponentially with $N$---so as the number of vertices $N$ increases, so AMI approaches $0$ superexponentially; for even small networks, $1/\mathcal{B}_N \approx 0$. In this way, AMI provides a ``geometry-free'' space in which no one partition is {\it a priori} closer to all others. This key property of AMI, called homogeneity, is proved in a Lemma in the next section. 

\begin{table}[t]
 \caption{Normalized mutual information for partitions in Fig.~\ref{fig_partitions}}
 \label{tab_nmi}
 \begin{center}
 \begin{tabular}{c|ccccc} 
 \hline
 & \multicolumn{5}{c}{Partition 2} \\ 
 Partition 1 & \textbf{1} & \textbf{2} & \textbf{3} & \textbf{4} & \textbf{5} \\
 \hline
 \textbf{1} & 1 & 0 & 0 & 0 & 0 \\
 \textbf{2} & 0 & 1 & 0.27 & 0.27 & 0.76 \\
 \textbf{3} & 0 & 0.27 & 1 & 0.27 & 0.76 \\
 \textbf{4} & 0 & 0.27 & 0.27 & 1 & 0.76 \\
 \textbf{5} & 0 & 0.76 & 0.76 & 0.76 & 1 \\
 \hline 
 $\mathbb{E}$[NMI] & 0.20 & 0.46 & 0.46 & 0.46 & 0.66 \\
 \hline
 \end{tabular}
 \end{center}
\end{table}

\begin{table}[t]
 \caption{Adjusted mutual information for partitions in Fig.~\ref{fig_partitions}}
 \label{tab_ami}
 \begin{center}
 \begin{tabular}{c|ccccc} 
 \hline
 & \multicolumn{5}{c}{Partition 2} \\ 
 Partition 1 & \textbf{1} & \textbf{2} & \textbf{3} & \textbf{4} & \textbf{5} \\
 \hline
 \textbf{1} & 1 & 0 & 0 & 0 & 0 \\
 \textbf{2} & 0 & 1 & -0.5 & -0.5 & 0 \\
 \textbf{3} & 0 & -0.5 & 1 & -0.5 & 0 \\
 \textbf{4} & 0 & -0.5 & -0.5 & 1 & 0 \\
 \textbf{5} & 0 & 0 & 0 & 0 & 1 \\
 \hline 
 $\mathbb{E}$[AMI] & 0.20 & 0 & 0 & 0 & 0.20 \\
 \hline
 \end{tabular}
 \end{center}
\end{table}

\subsubsection{Lemma and theorems} We now prove a lemma about adjusted mutual information, and then formally state the NFL theorem for supervised learning and prove the no free lunch theorem for community detection.

\vspace{0.1in}
\begin{addmargin}[3em]{3em}
\noindent{\bf Lemma 1}: Adjusted mutual information (AMI) is a homogenous loss function over the interior of the space of partitions of $N$ objects. Including the boundary partitions, i.e., the $1$-partition and the $N$-partition, AMI is homogenous within $\mathcal{B}_N^{-1}$.
\end{addmargin}

\vspace{0.05in}
\noindent{\it Proof:} Showing that AMI is a homogenous loss function requires that we show 
\begin{equation}
	L(u) = \sum_{v \in \Omega} \text{AMI}(u,v)
	\label{wts}
\end{equation}
is independent of $u$, where $\Omega$ is the space of all partitions of $N$ objects. Stated plainly, if $L(u)$ is independent of $u$, it means that the total AMI between partition $u$ and all possible partitions will be the same, no matter which partition $u$ is chosen. The definition of AMI is: 
$$\text{AMI}(u,v) = \frac{I(u,v) - E[I(u,v)]}{\sqrt{H(u)H(v)} - E[I(u,v)]}$$
where $I$ is mutual information and $H$ is entropy \cite{vinh2009information}. The AMI takes on a value of $1$ when two partitions are identical and a value of $0$ when they are only correlated to the extent that one would expect by chance. In particular, the expectation $E$ is taken over all possible pairs of partitions $u'$ and $v'$ such that every $u'$ has the same number of groups and the same number of objects belonging to each group as does $u$, and likewise for $v'$ and $v$. In this way, the expectation $E$ is taken over all pairs of divisions that preserve the group sizes of the two partitions being compared. For convenience of notation, let $\phi$ be a subset of all partitions $\Omega$ such that every partition $v \in \phi$ has the same number of groups and same number of objects in each group. The set of all partitions $\Omega$ may be subdivided into non-overlapping subsets $\{ \phi_i\}$, such that $\cup_i \phi_i = \Omega$ and $\phi_i \cap \phi_j = \varnothing$ for any $i \neq j$. (For example, in Fig.~\ref{fig_partitions}, partition $1$ belongs to $\phi_1$, partitions $2$, $3$, and $4$ belong to $\phi_2$, and partition $5$ belongs to $\phi_3$.) Let the particular subset $\phi_i$ to which a partition $u$ belongs be denoted by $\phi(u)$.

Prior to proceeding, we note that there are two special boundary partitions, the $1$-partition in which all objects are in a single group and the $N$-partition in which each object is in its own group. These will be denoted by $\bar 1$ and $\bar N$ respectively. Note that $\bar 1 = \phi(\bar 1)$ so that $\lvert \phi(\bar 1) \rvert = 1$, and that $\phi(\bar N)$ is equivalent to the set of all possible relabelings of the $N$ objects, so that $\lvert \phi(\bar N) \rvert = N!\ $. Because there is only one element of $\phi(\bar 1)$, it is necessarily true that $I(\bar 1, \bar 1) = E[I(\bar 1, \bar 1)] = H(\bar 1)$. Thus, for this special case, the numerator and denominator of AMI are identical, and $\text{AMI}(\bar 1,\bar 1) = 1$. Similarly, because the set $\phi(\bar N)$ contains every possible permutation of the labels of the objects, yet all partitions are identical, $I(\bar N, \bar N) = E[I(\bar N, \bar N)] = H(\bar N)$, and so $\text{AMI}(\bar N, \bar N) = 1$. 

In order to prove Eq.~\eqref{wts}, we will show that $L(u)=0$ for all $u$ except $\bar 1$ and $\bar N$ by demonstrating that the numerator of the definition of AMI is 0, specifically,
\begin{equation}
	\sum_{v \in \Omega} \left [ I(u,v) - E[I(u,v)] \right ]= 0 \ \ \forall\ u \neq \bar 1 \text{ or } \bar N\ .
	\label{wts2}
\end{equation}
In fact, we will show that Eq.~\eqref{wts2} holds by breaking the entire sum over all partitions $\Omega$ into sums over each of its disjoint subsets $\{\phi_i\}$, and proving that  
\begin{align}
	&\sum_{v' \in \phi(v)} \left [ I(u, v') - E[I(u,v')] \right ]= 0 \nonumber \\ 
	&\forall\ u \text{ and }\forall\ v \text{ except } u=v=\bar 1 \text{ or } u = v = \bar N\ .
	\label{wts3}
\end{align}
In other words, we will show that the numerator of the definition of AMI is equal to zero when summed over any subset $\phi(v)$ for any fixed partition $u$, except the boundary cases that both $u$ and $v$ are equal to $\bar 1$ or both are equal to $\bar N$. We first examine the expectation term in Eq.~\eqref{wts3}. Recall that the expectation is taken over all pairs of members of the subsets $\phi(u)$ and $\phi(v)$, respectively,
\begin{equation}
	E[I(u,v)] = \frac{1}{|\phi(u)||\phi(v)|} \sum_{u' \in \phi(u)} \sum_{v' \in \phi(v)} I(u',v')\ .
	\label{expect}
\end{equation}
In fact, because the sums above are taken over the subsets $\phi(u)$ and $\phi(v)$ that contain $u$ and $v$, the expected mutual information is equal to a constant $\zeta$ for any pair of partitions drawn from $\phi(u)$ and $\phi(v)$, 
\begin{equation}
	E[I(u,v)] = 	\zeta\quad \forall\ u\in\phi(u) \text{ and } \forall\ v\in\phi(v)\ .
	\label{equiv}
\end{equation}
Note then that we may rewrite the sum over expectations in Eq.~\eqref{wts3} as $\sum_{v' \in \phi(v)} E[I(u,v')] = \lvert \phi(v)\rvert\  \zeta$. Therefore, it remains to be shown that the sum over mutual informations in Eq.~\eqref{wts3} is also equal to $\lvert \phi(v)\rvert\  \zeta$, 
\begin{equation}
	\sum_{v' \in \phi(v)} I(u, v ') = \lvert \phi(v)\rvert\  \zeta\ .
	\label{xfixed}
\end{equation}
To see that this is true, despite the fact that $u$ is fixed (and not averaged over all $u' \in \phi(u)$ as in $E[I(u,v)]$), note that Eq.~\eqref{xfixed} nevertheless sums over every $v' \in \phi(v)$ which is the set of every randomization of the partition $v$, provided group sizes are held constant. Because this includes all relabelings (or reindexings) of the $N$ objects being partitioned, it must be true that,
\begin{equation}
	\sum_{v' \in \phi(v)} I(u_1, v ') = \sum_{v' \in \phi(v)} I(u_2, v ') \text{ whenever } u_1 \in \phi(u_2)\ .
\end{equation}
In other words, the sum of mutual information between a fixed partition $u_1$ and all members of a subset $\phi(v)$ must be equal to the sum of mutual information between a different fixed partition $u_2$ and the same subset $\phi(v)$, but only if $u_1$ and $u_2$ both belong to the same subset as each other. Therefore, Eq.~\eqref{xfixed} is true, meaning that the sum over the two terms in Eq.~\eqref{wts3} is zero, independent of $u$. This first implies that the AMI between any boundary partition and any interior partition is $0$, $\text{AMI}(u,\bar 1) = 0$ for any $u \neq \bar 1$ and $\text{AMI}(u,\bar N) = 0$ for any $u \neq \bar N$. This, in turn, implies Eq.~\eqref{wts2} is true. This completes the proof of the first statement, that Eq.~\eqref{wts} is true, and in particular, $L(u) = 0$, for any $u\neq \bar 1, \bar N$ and AMI is homogeneous over all non-boundary partitions.  

In the special cases of $u=v=\bar 1$ and $u=v=\bar N$, note that we have already shown that $\text{AMI}(\bar 1, \bar 1) = 1$, $\text{AMI}(\bar N, \bar N) = 1$, and $\text{AMI}(u,\bar 1) = 0$ for any $u \neq \bar 1$ and $\text{AMI}(u,\bar N) = 0$ for any $u \neq \bar N$. Therefore, 
\begin{align}
	L(\bar 1) = \sum_{v \in \Omega} \text{AMI}(\bar 1,v) = \mathcal{B}_{N}^{-1}\ , \nonumber \\
	L(\bar N) = \sum_{v \in \Omega} \text{AMI}(\bar N,v) = \mathcal{B}_{N}^{-1}\ ,
\end{align}
completing the proof of the second statement: including the boundary points, AMI is homogenous within an additive constant $\mathcal{B}_{N}^{-1}$. \qed

\vspace{0.1in}
\begin{addmargin}[3em]{3em}
\noindent{\bf Theorem 2 (Wolpert 1996)}: For homogeneous loss $\ell$, the uniform average over all $\gamma$ of $P(c | \gamma,d)$ equals $\Lambda(c) / r$.
\end{addmargin}

\vspace{0.05in}
\noindent{\it Proof:} See Ref.~\cite{wolpert1996lack}.

\vspace{0.1in}
\begin{addmargin}[3em]{3em}
\noindent{\bf Theorem 3 (No free lunch for community detection)}: For the community detection problem with a loss function of adjusted mutual information, the uniform average over all $\gamma$ of $P(c | \gamma)$ equals $\Lambda(c) / r$.
\end{addmargin}

\vspace{0.05in}
\noindent{\it Proof:} Lemma 1 proves that adjusted mutual information is homogeneous and applying Theorem 2 with $d=\varnothing$ completes the proof. \qed

\subsubsection{Implications}

No free lunch for community detection means that, uniformly averaged over all community detection problems, and evaluated by AMI, all algorithms have equivalent performance. Phrased more usefully, it means that any subset of problems for which an algorithm outperforms others is balanced by another subset for which the algorithm underperforms others. Thus, there is no single community detection algorithm that is best overall. 

On the other hand, if the set of problems of interest is a non-uniform subset of all problems, then one algorithm may outperform another on this subset. In other words, the bias of an algorithm to solving a particular type of community detection problem may be its strength, accepting the fact that such an advantage must be balanced by disadvantages elsewhere. 
For instance, algorithms like the unconstrained SBM (which can find both assortative and disassortative communities and mixtures and gradations thereof) are not universally superior to versions of the SBM constrained to find only assortative or disassortative communities~\cite{larremore2014efficiently}---if the particular subset of problems is believed to contain only disassortative communities, then the unconstrained SBM will not perform as well as a constrained one. In other words, no free lunch for community detection means that matching the assumptions in the model to the underlying generative process can lead to better, more accurate results, but only in the cases when the beliefs about the underlying generative process are accurate; in the other cases, the same model assumptions that improved performance on some problems will diminish it for others. 
To some extent we expect the distribution of problems to be non-uniform in general. Out of all the possible ways of constructing a graph there may be some types of graph we are less likely to observe. For each graph we can also expect that of all the possible partitions, many will correspond to random assignments of nodes that are not useful in any application. Put differently, there may be some problems we do not wish to solve---but, unless we know which problems they are, it offers us little or no benefit in practice. We note that relatively little is known about which algorithms perform better than others within particular domains or on particular classes of networks. A valuable line of future research on community detection will be developing such an understanding~\cite{peel2011estimating, yang2016comparative}. 

\noindent\matrixquote{{\bf Cypher}: \textit{You know, I know this steak doesn't exist. I know that when I put it in my mouth, the Matrix is telling my brain that it is juicy and delicious. After nine years, you know what I realize? $\dots$ Ignorance is bliss.}~\cite{matrix1999}}

\clearpage
\section{Datasets and additional methodology}\label{supp:data}

\noindent\matrixquote{\textbf{Neo}:\textit{\href{https://www.youtube.com/watch?v=RP8uhXuS2n8}{Whoa.}}~\cite{matrix1999}}

\vspace{-0.3in}
\subsection{Lazega Lawyers networks}
The Lazega Lawyers network is a multilayer network consisting of $71$ attorneys of a law firm with three different sets of links, corresponding to friendships, exchange of professional advice, and shared cases~\cite{lazega2001collegial}. The original study also collected five sets of categorical node metadata, corresponding to status (partner or associate), gender, office location, type of practice (corporate or litigation), and law school (Harvard, Yale, UConn, other). The relationships and dynamics within the law firm were studied extensively in the initial publication of these data sets, but they were not primarily analyzed as complex networks. 

\subsection{Malaria {\it var} gene networks}
The Malaria data set consists of $307$ {\it var} gene sequences from the malaria parasite {\it P. falciparum} \cite{larremore2013network}. Each {\it var} gene encodes a protein that the parasite uses to evade the human immune system, and therefore this family of genes is under intense evolutionary pressures from the human host. The original study focused on uncovering the functional and evolutionary constraints on {\it var} gene evolution by identifying community structure in {\it var} gene networks. 

These sequences were independently analyzed at $9$ loci (locations within the genes), producing $9$ different genetic-substring-sharing networks with the same node set. In other words, there are $9$ layers in this multilayer network. Each parasite genome contains around $60$ {\it var} genes, and the $307$ genes in this data set represent seven parasite genomes. The original study included three sets of categorical node metadata, corresponding to the upstream promoter sequence classification (UPS, $K=3$), CysPoLV groups (CP $K=6$), and the parasite genome from which sequence was generated (parasite origin $K=7$). 

\subsection{Karate Club network}
The Zachary Karate Club represents the observed social interactions of $34$ members of a karate club~\cite{zachary1977information}. At the time of study, the club fell into a political dispute and split into two factions, which are treated as metadata. The Karate Club has been analyzed exhaustively in studies of community detection, and its faction metadata have often been used as ground truth for community detection, due to the network's small size and easily interpretable social narrative. 

\subsection{Generation of log-likelihood surface plots}
\label{sec_surfplots}

The log-likelihood surface plots in Figs.~\ref{fig_karate_optima}, \ref{fig_synth_neo}, \ref{fig_laz_neo}, \ref{fig_karate_neo}, \ref{fig:malaria_neo_1}, and \ref{fig:malaria_neo_6} illustrate the changes in log likelihood as the partition of network nodes is varied. In the figures, we show surfaces that appear to be continuous over that two dimensional space, in spite of the fact that the true space of partitions is high dimensional and discretized, and so here we explain the methods used to produce visually meaningful plots.

Plots were generated in three steps: partition sampling, data projection and surface interpolation. For most networks it is infeasible to calculate the log likelihood of all possible partitions, so we instead sampled a subset of partitions. We began with the set of partitions along the path of the neoSBM (e.g.,~Fig.~\ref{fig_synth_neo}) and sampled partitions around the local neighborhood of this initial set. Specifically, we did so by selecting two partitions uniformly at random from the initial set and created each new partition by assigning $q$ nodes (chosen randomly and uniformly) to the group assignment of the first partition and the remaining $N-q$ nodes to that of the second partition.

Next, we projected the $K^N$-dimensional partition data down to two dimensions using Multi-dimensional Scaling (MDS)~\cite{borg2005modern} and variation of information~\cite{meila2003comparing} as a similarity measure.  The outcome of this projection was a two-dimensional representation of the partition space that preserves the variation of information between partitions.

Finally, we used MATLAB's \textit{scatteredInterpolant} function with \textit{natural} interpolation to fit an interpolated surface to the data, which we evaluated over a grid of domain points and smoothed using a Gaussian kernel to improve legibility. The processes of embedding, interpolating, and smoothing are not particularly sensitive to changes in parameters or grid resolutions.

In the special case of Fig.~\ref{fig_laz_neo}, we also plotted the partitions of the neoSBM in addition to the interpolated log-likelihood surface to illustrate the neoSBM's path in the broader context of the surface. There were no modifications or smoothing of the points of the neoSBM's path beyond the embedding process described above.

\noindent\matrixquote{\textbf{Neo}:\textit{ I know you're out there. I can feel you now. I know that you're afraid $\dots$ you're afraid of us. You're afraid of change. I don't know the future. I didn't come here to tell you how this is going to end. I came here to tell you how it's going to begin. I'm going to hang up this phone, and then I'm going to show these people what you don't want them to see. I'm going to show them a world without you. A world without rules and controls, without borders or boundaries. A world where anything is possible. Where we go from there is a choice I leave to you.}~\cite{matrix1999}}

\end{document}